\documentclass[preprint, superscriptaddress]{revtex4-2}

\usepackage{epstopdf}
\usepackage{graphicx}
\usepackage{dcolumn}
\usepackage{multirow}
\usepackage{bm}
\usepackage{amsthm,amsmath,amssymb}
\usepackage{mathrsfs}

\usepackage[colorlinks=true,citecolor=blue]{hyperref}
\usepackage{soul}

\usepackage{titlesec}

\begin{document}

\title{Deterministic and Efficient Switching of Sliding Ferroelectrics}

\author{Shihan Deng}
\thanks{These authors contributed equally.}
\affiliation{Key Laboratory of Computational Physical Sciences (Ministry of Education), Institute of Computational Physical Sciences, State Key Laboratory of Surface Physics, and Department of Physics, Fudan University, Shanghai 200433, China}
\affiliation{Shanghai Qi Zhi Institute, Shanghai 200030, China}

\author{Hongyu Yu}
\thanks{These authors contributed equally.}
\affiliation{Key Laboratory of Computational Physical Sciences (Ministry of Education), Institute of Computational Physical Sciences, State Key Laboratory of Surface Physics, and Department of Physics, Fudan University, Shanghai 200433, China}
\affiliation{Shanghai Qi Zhi Institute, Shanghai 200030, China}

\author{Junyi Ji}
\affiliation{Key Laboratory of Computational Physical Sciences (Ministry of Education), Institute of Computational Physical Sciences, State Key Laboratory of Surface Physics, and Department of Physics, Fudan University, Shanghai 200433, China}
\affiliation{Shanghai Qi Zhi Institute, Shanghai 200030, China}

\author{Changsong Xu}
\email[]{csxu@fudan.edu.cn}
\affiliation{Key Laboratory of Computational Physical Sciences (Ministry of Education), Institute of Computational Physical Sciences, State Key Laboratory of Surface Physics, and Department of Physics, Fudan University, Shanghai 200433, China}
\affiliation{Shanghai Qi Zhi Institute, Shanghai 200030, China}

\author{Hongjun Xiang}
\email[]{hxiang@fudan.edu.cn}
\affiliation{Key Laboratory of Computational Physical Sciences (Ministry of Education), Institute of Computational Physical Sciences, State Key Laboratory of Surface Physics, and Department of Physics, Fudan University, Shanghai 200433, China}
\affiliation{Shanghai Qi Zhi Institute, Shanghai 200030, China}

\date{\today}

\begin{abstract}
    Recent studies highlight the scientific importance and broad application prospects of two-dimensional (2D)  sliding ferroelectrics, which prevalently exhibit vertical polarization with suitable stackings. It is crucial to understand the mechanisms of sliding ferroelectricity and to deterministically and efficiently switch the polarization with optimized electric fields. Here, applying our newly developed DREAM-Allegro multi-task equivariant neural network, which simultaneously predicts interatomic potentials and Born effective charges, we construct a comprehensive potential for boron nitride ($\mathrm{BN}$) bilayer. The molecular dynamics simulations reveal a remarkably high Curie temperature of up to 1500K, facilitated by robust intralayer chemical bonds and delicate interlayer van der Waals(vdW) interactions. More importantly, it is found that, compared to the out-of-plane electric field, the inclined field not only leads to deterministic switching of electric polarization, but also largely lower the critical strength of field,  due to the presence of the in-plane polarization in the transition state. This strategy of an inclined field is demonstrated to be universal for other sliding ferroelectric systems with monolayer structures belonging to the symmetry group $p \bar{6} m 2$, such as transition metal dichalcogenides (TMDs).
\end{abstract}

\maketitle

    Achieving ferroelectric devices with small thickness and perpendicular polarization is a critical step toward realizing low-energy, nonvolatile, and high-density ferroelectric memory \cite{Yasuda2021}. However, it is well-known that the depolarization field becomes significant in thin film ferroelectrics. Additionally, ``dead layers'' inevitably form due to factors such as interface effects, defects, and impurities, all of which lead to performance degradation in the application of ultrathin films \cite{Stern2021, Müller2015, Wang2023, Stengel2006}. 

\begin{figure}[t]
        \includegraphics[width=8cm]{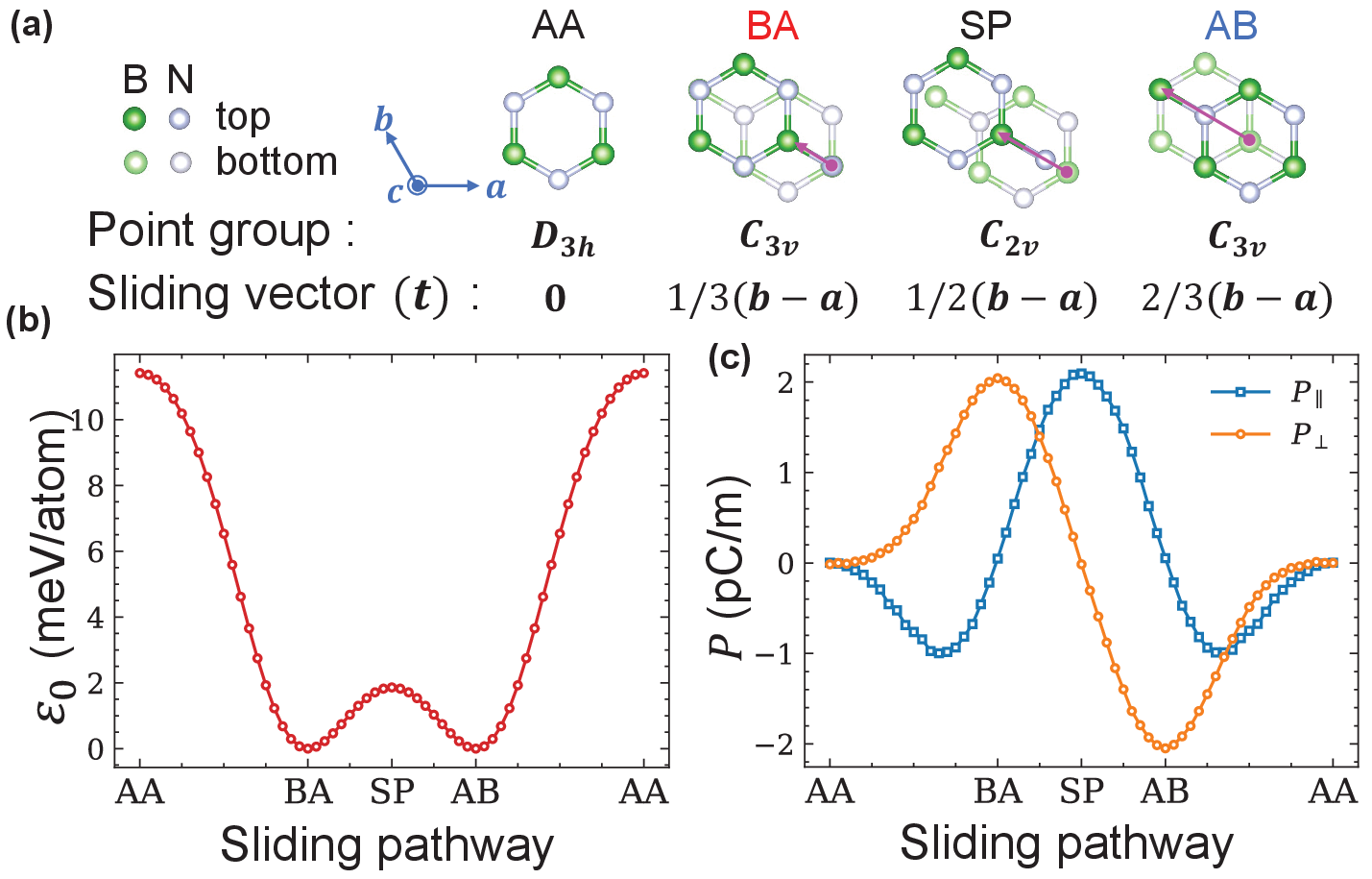}
        \caption{\label{fig:1-Stacking} (a) Schemes of the AA, BA, AB, and SP stackings of $\mathrm{BN}$ bilayers. Blue arrows define the lattice vectors $\bm{a}$, $\bm{b}$ and $\bm{c}$ of unit cell. Purple arrows indicate the sliding vector, $\bm{t}$, which is defined as the fractional displacement of the top layers with respect to bottom layer in the AA stacking. Energy $\varepsilon_0$ (b) and polarization $P$ (c) along the designed sliding pathway, as calculated by the DFT and Berry phase approach \cite{KingSmith1993, Resta1992}.}
    \end{figure}

    The recently proposed concept of sliding ferroelectricity in 2D vdW materials naturally overcomes these obstacles \cite{LiLei2017}. The prototype of sliding ferroelectrics is $\mathrm{BN}$ bilayers [see Fig.~\ref{fig:1-Stacking}(a)], which display energy ground state as the AB (BA) configuration \cite{Ribeiro2011, Stern2021, Yasuda2021, WuMenghao2021_2,Constantinescu2013, Evgeny2021} with $C_{3v}$ symmetry. Within the AB configuration, the top $\mathrm{B}$ atom aligns with the bottom $\mathrm{N}$ atom, and the top $\mathrm{N}$ atom and the bottom $\mathrm{B}$ atom sit at the hollow sites of the honeycomb lattice. The AB(BA) stacking mode disrupts mirror symmetry and leads to an out-of-plane polarization of 2 pC/m in the $-z$($z$) direction \cite{LiLei2017,JiangWen2022}. This vertical polarization arises from charge redistribution driven by interlayer coupling \cite{Stern2021, ZhengZhiren2020, Wang2023, LiuMeng2023}. The BA state can transition to the AB state via relative sliding between the two layers by a $\mathrm{B}$-$\mathrm{N}$ bond length in any of the three rotationally symmetric directions \cite{JiangWen2022,LiuMeng2023,TangPing2023}. Conversely, sliding in the opposite direction by one bond length leads to the mirror-symmetric ($D_{3h}$) AA state, representing the highest energy state. Sliding ferroelectrics, especially the proposal of bilayer stacking ferroelectricity (BSF) theory \cite{JiJunyi2023}, broaden the spectrum of candidate materials for 2D ferroelectrics, as bilayers made of nonpolar monolayers can exhibit ferroelectricity through specific stackings. Beyond $\mathrm{BN}$ bilayer, sliding ferroelectrics has been experimentally confirmed in various vdW systems, including semimetals \cite{FeiZaiyao2018, Sharma2019, XiaoJun2020, Barrera2021, Jindal2023}, insulators \cite{LvMing2022}, semiconductors \cite{HuHaowen2019, SuiFengrui2023, WangXirui2022, SunYan2022, WanYi2022}, and organic crystals \cite{MiaoLeping2022}, all of which are robust at room temperature. Notably, recent experiments highlighting high endurance and fatigue resistance in sliding ferroelectrics underscore the significant potential of these materials for practical applications \cite{Yasuda2024, BianRenji2024}.

    The study of sliding ferroelectricity is rapidly developing, with both fundamental and practical challenges yet to be addressed. Although the switching energy barriers are typically very low ($\sim$meV per unit cell \cite{LiLei2017, YangQing2018, WuMenghao2021}), sliding ferroelectrics exhibit high Curie temperatures ($T_C$), as most of them are operated at room-temperature. Thermodynamic modeling using a mean-field approximation suggests $T_C=1.58 \times 10^4$ K for $\mathrm{BN}$ bilayer \cite{TangPing2023}. Such prediction reflects the good stability of $\mathrm{BN}$ bilayers. Moreover, the polarization of sliding ferroelectrics is typically small, leading to a large coercive field for polarization switching, which indicates that a wiser and more efficient switching strategy is highly desired. Symmetry changes during slidings indicating that the Born effective charges (BECs) can change dramatically \cite{Bennett2024}, the treatment of fixed BEC may be insufficient \cite{HeRi2024} [see Fig.S1 in the supplemental material (SM) \cite{SM} for further discussion\nocite{Kresse1996, Blochl1994, Perdew1996, Grimme2006, JiangWen2022, Vaspkit2019, Jinnouchi2019, Jinnouchi2019_2, Jinnouchi2020, Allen2017, Gonze1997, Musaelian2023, YuHongyu2024, LAMMPS2022, Nosé1984, origin2024, WangXirui2022, Weston2020}]. Ab initio molecular dynamics (AIMD) can be used to study sliding ferroelectricity, but it cannot handle large systems and is computationally consuming \cite{ShiBowen2023}. Hence, an accurate and efficient description of both intralayer and interlayer couplings is required.

    In this Letter, applying our newly developed DREAM method (i.e., a generalized dielectric response equivariant atomistic multi-task framework based on Allegro \cite{Musaelian2023}), we construct a neural network model for $\mathrm{BN}$ bilayer. Such model is capable of predicting both the interatomic potential and the BEC tensors, which enable prediction of a reasonable $T_C$ and accurate responses to applied electric fields. Moreover, it is found that an inclined electric field, which breaks the three-fold rotational symmetry of $\mathrm{BN}$ bilayer, can not only deterministically control directions of slidings, but also substantially reduce the total coercive field. These findings are general and can be applied to other similar sliding ferroelectrics.

    \textcolor{blue}{Ferroelectric transition of BN bilayer.} 
    Firstly, the basic properties of $\mathrm{BN}$ bilayers are examined by density functional theory (DFT) calculations. As illustrated in Fig.~\ref{fig:1-Stacking}(b)(c), the out-of-plane polarization ($P_{\perp}$) of the AB (BA) state yields $\mp$2.078 pC/m, and the energy barrier for polarization switching is determined to be 7.44 meV per unit cell. Such results are well consistent with those of previous studies \cite{LiLei2017, JiangWen2022}. Notably, the transition state or intermediate saddle-point (SP) state with $C_{2v}$ symmetry exhibits a strictly in-plane polarization ($P_{\parallel}$) of 2.091 pC/m. 

    To train the potential with the original Allegro method \cite{Musaelian2023}, we run on-the-fly machine learning force fields(MLFFs) \cite{Jinnouchi2019, Jinnouchi2019_2, Jinnouchi2020} and generate 1659 structures with $5\times5\times1$ supercell, across a wide temperature range below 3000 K. The structure, energy, force, and stress tensor are included in training. The obtained model can accurately predict the energy difference among structural configurations, with a small mean absolute errors (MAEs) of 0.053 meV/atom [see SM \cite{SM} for details].

    \begin{figure}[t]
        \includegraphics[width=7cm]{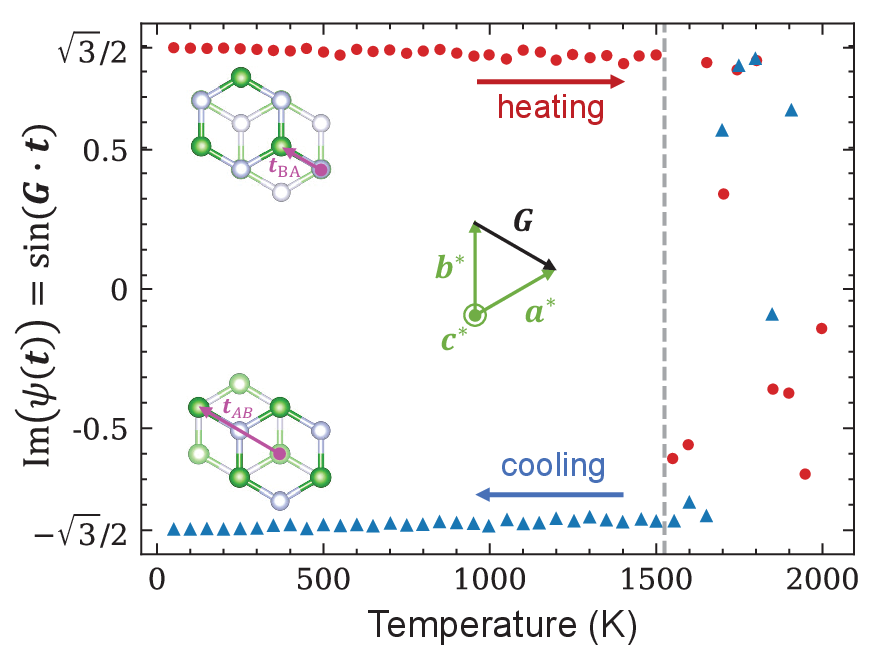}
        \caption{\label{fig:2-Tc} The change of $\mathrm{Im}(\psi(\bm{t}))$ as a function of temperature. Red dots and blue triangles represent the heating and cooling processes, respectively, which are marked by  red and blue arrows. Green arrows denote the reciprocal lattice vectors $\bm{a}^*$ and $\bm{b}^*$, and the black arrow indicates that $\bm{G} = \bm{a}^*-\bm{b}^*$ is selected. Purple arrows visualize the sliding vector $\bm{t}$ of the BA and AB state, which are $\bm{t}_{\mathrm{BA}}=1/3(\bm{b}-\bm{a})$ and $\bm{t}_{\mathrm{AB}}=2/3(\bm{b}-\bm{a})$ respectively. The dashed gray line signifies the critical temperature.}
    \end{figure}

    In the case of sliding ferroelectricity, $T_C$ can be characterized by the sliding vector $\bm{t}$ \cite{TangPing2023}:
    \begin{equation}
        \lim\limits_{T \to T_C^-} \frac{\partial \langle \bm{t} \rangle} {\partial T} \to \infty,
    \end{equation}
    where $\langle~\rangle$ signifies the ensemble average at a certain temperature. Vector $\bm{t}$ is the averaged displacements of the top layer with respect to the bottom layer in the AA stacking [see Fig.~\ref{fig:1-Stacking}(a)]. To avoid the impact of in-plane periodic boundary conditions, we use the reciprocal lattice vector ($\bm{G}$) to define a new order parameter ($\psi$), which reads
    \begin{equation}
        \psi(\bm{t}) = e^{i\bm{G}\cdot\bm{t}}.
    \end{equation}
    where $\psi$ ranges between $[-1,1]$ and remains unchanged under the substitution $\bm{t} \to \bm{t} + \bm{R}$, where $\bm{R}$ represents the in-plane lattice vector.
    
    We now focus on the ferroelectric-paraelectric transition of $\mathrm{BN}$ bilayer. The MD simulations are performed using our potential model, where 900 atoms is incrementally heated and cooled between 1 K and 2000 K, with an interval of 50 K. At each temperature, the quantity of $\psi$ is computed for the equilibrium states over 50 ps. The changes in the imaginary part of $\psi$ as a function of temperature are presented in Fig.~\ref{fig:2-Tc} for $\bm{G} = \bm{a}^* - \bm{b}^*$. Starting in the BA state [$\mathrm{Im}(\psi_{\mathrm{BA}})= \sqrt{3}/2$], an increase in temperature progressively triggers system sliding. Conversely, during the cooling process, system sliding gradually ceases and eventually randomly stabilizes in the AB state [$\mathrm{Im}(\psi_{\mathrm{AB}})= -\sqrt{3}/2$]. Accoriding to Fig.~\ref{fig:2-Tc}, the critical temperature is approximated as 1500 K. Intriguingly, even for temperatures over $T_C$, the system conducts only few complete slides within 50 ps, oscillating around the BA or AB state most of the time. This phenomenon causes $\psi$ to oscillate above and below zero, contrasting with traditional order parameters that promptly vanish above $T_C$. The presently predicted $T_C=1500$ K is much more reasonable than the $1.58 \times 10^4$ K from mean-field approximation \cite{TangPing2023}.

    \textcolor{blue}{Switching of sliding ferroelectric $\mathrm{BN}$ bilayer.}
    To simulate the $\mathrm{BN}$ bilayer under an external electric field, we integrate Allegro within the new DREAM framework \cite{YuHongyu2024, SM}. This integration enables the network to learn both scalars and tensors by leveraging geometric tensors \cite{Batzner2022, Gasteiger2022}. Consequently, the network is capable of simultaneously predicting interatomic potentials and BECs. The MAEs of the final model, which incorporates BEC information of 3484 structures at temperatures below 500K, yields 0.030 meV/atom in energy and 0.008 $|$e$|$ in BEC, demonstrating good accuracy of the model.

    \begin{figure}
        \includegraphics[width=8cm]{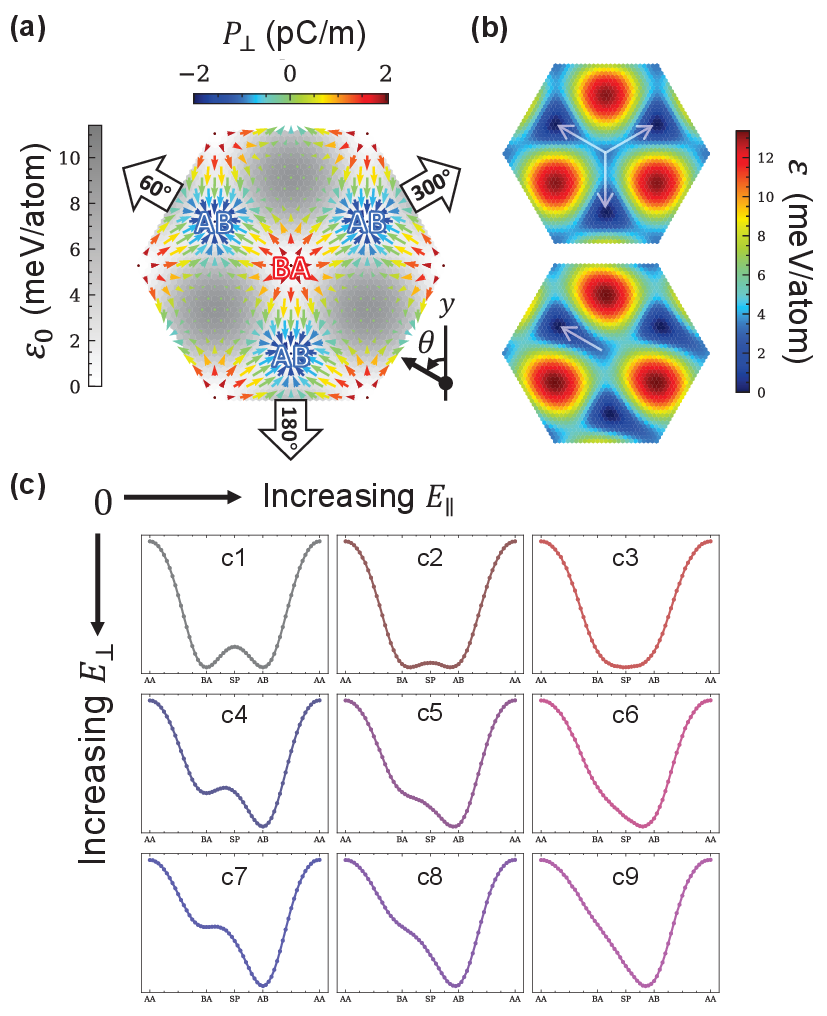}
        \caption{\label{fig:3-E_analytical} (a) Energy $\varepsilon_0$ and polarization $P$ distribution calculated by the DFT and Berry phase approach. Positions within the hexagon correspond to different sliding vectors, with the center representing the BA state. The shades of grey in the background indicate energy levels. The direction of colored arrows shows in-plane polarizations $P_{\parallel}$, while their colors represent out-of-plane polarization $P_{\perp}$. The sliding directions $\theta$ are indicated by large hollow arrows. (b) The energy surface with only $E_{\perp}$ along $-z$ (the top panel) and with both $E_{\perp}$ along $-z$ and $E_{\parallel}$ at $60^{\circ}$ (the bottom panel), calculated according to Eq.~\ref{eqn_Energy} with $E_{\perp}$ and $E_{\parallel}$ equal to 1V/{\AA}. White arrows denote the sliding pathways with the lowest barrier. (c) Energy along the sliding pathway under varying $E_{\parallel}$ and $E_{\perp}$, also calculated using Eq.~\ref{eqn_Energy}. The $E_{\parallel}$ increases from left to right panels, while $E_{\perp}$ are gradually enhanced from top to bottom panels.}
    \end{figure}

    We now examine the responses of the $\mathrm{BN}$ bilayer to perpendicular field $E_{\perp}$. We start with the ferroelectric BA state and incrementally increase $E_{\perp}$ along the $-z$ direction over time at a rate of 0.02 V/({\AA}$\cdot$ps), at temperatures ranging from 100 K to 500 K. The sliding process occurs rapidly, typically within a few picoseconds, and thermal vibrations add complexity to identifying the exact onset of sliding. Therefore, we define the coercive field ($E_{\perp, c}$) as the field at which sliding reaches the SP state. Our results indicate that $E_{\perp, c}$ decreases with increasing temperature, starting at approximately 1.99 V/{\AA} at 100 K and saturating at around 1.39 V/{\AA} for over 200 K. Moreover, since $E_{\perp}$ does not break the three-fold rotational symmetry of the $\mathrm{BN}$ bilayer, sliding occurs randomly along the three rotationally symmetric directions in each simulation [see the top panel of Fig.~\ref{fig:3-E_analytical}(b)].

    Given that some sliding states of $\mathrm{BN}$ bilayer exhibit $P_{\parallel}$, we investigate the effect of in-plane electric field ($E_{\parallel}$) on ferroelectric switching. As revealed by Fig.~\ref{fig:3-E_analytical}(a), the direction of $P_{\parallel}$ is the same as the corresponding sliding direction. These directions can be characterized by the angle $\theta$, which is measured through a counterclockwise rotation from the $y$-axis. This finding implies that $E_{\parallel}$ oriented along one of these three directions can break the three-fold rotational symmetry and induce deterministic switching. For instance, $E_{\parallel}$ oriented at $60^{\circ}$ will lead to sliding being only along $60^{\circ}$ [see the lower panel of Fig.~\ref{fig:3-E_analytical}(b)]. Actually, applying $E_{\parallel}$ within the range of $(0^{\circ},120^{\circ})$ can all leads to sliding along $60^{\circ}$ [refer to Fig.S4]. In addition, Fig.~\ref{fig:3-E_analytical}(a) shows significant $P_{\parallel}$ at the bridging SP state, which indicates that suitable $E_{\parallel}$ can substantially reduces the energy barrier. Such conjecture can be verified by energy distribution shown in the first two columns of Fig.~\ref{fig:3-E_analytical}(c)], where the energy barrier for sliding from BA to AB state is obviously reduced in the presence of $E_{\parallel}$.

    \begin{figure}[t]
        \includegraphics[width=8cm]{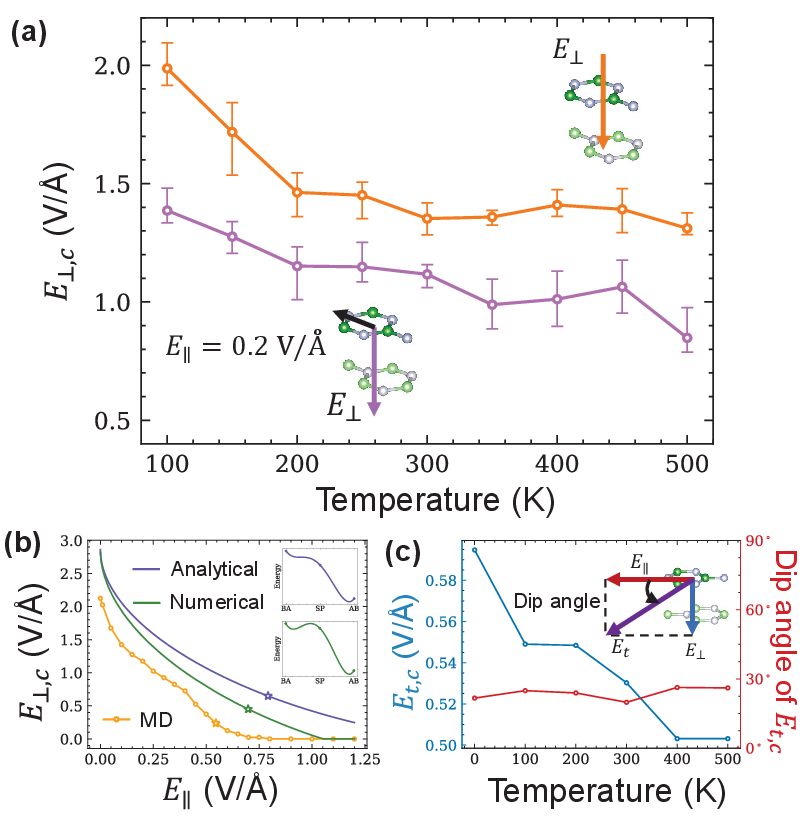}
        \caption{\label{fig:4-E_MD} (a) $E_{\perp, c}$ as a function of temperature. The orange and purple lines represent the scenarios with $E_{\parallel} = 0$ and $E_{\parallel} = 0.2$V/{\AA}, respectively, at $\theta=60^{\circ}$. The error bars indicate the range of results from parallel simulations. (b) The relationship between $E_{\perp, c}$ and $E_{\parallel}$, as obtained by analytical calculations (purple line), numerical calculations (green line), and MD simulations at 0.1K (yellow line). The pentagram marks on the curves indicate the the minimum $E_{t, c}$. The insets depict the energy along the sliding pathway with the minimum $E_{t, c}$ obtained by analytical (the top panel) and numerical (the bottom panel) calculations, respectively. (c) The magnitude (blue line) and dip angle (red line) of the minimum $E_{t, c}$ across different temperatures. The dip angle is defined as the angle between $E_{t, c}$ and the horizontal plane.}
    \end{figure}

    Moreover, the decreasing in the energy barriers also indicate that the presence of $E_{\parallel}$ may reduce $E_{\perp, c}$. To examine such conjecture, we fit the energy $\varepsilon_0$, in-plane dipole $p_{\parallel}$, and out-of-plane dipole $p_{\perp}$ as a function of the amount of sliding $t$ with respect to the SP state, along the BA to AB sliding pathway. The fitted models yield $\varepsilon_0(t) = e_4 t^4 - 2e_4 t_{\mathrm{BA}}^2 t^2 + e_0$, $p_{\parallel}(t) = p_4^{\parallel} t^4 + p_2^{\parallel} t^2 - p_4^{\parallel} t_{\mathrm{BA}}^4 - p_2^{\parallel} t_{\mathrm{BA}}^2$, and $p_{\perp}(t) = p_3^{\perp} t^3 - 3p_3^{\perp} t_{\mathrm{BA}}^2 t$, where $t_{\mathrm{BA}}=-1/6|\bm{b}-\bm{a}|=-0.725$ {\AA} is a constant. The coefficients are determined to be $e_4=27.017$ meV/{\AA}$^4$, $e_0=7.41$ meV, $p_4^{\parallel}=5.82 \times 10^{-3}$ e/{\AA}$^3$, $p_2^{\parallel}=-1.66 \times 10^{-2}$ e/{\AA}, and $p_3^{\perp}=9.13 \times 10^{-3}$ e/{\AA}$^2$ [refer to Fig.S3]. The total potential energy after applying $E_{\parallel}$ along the BA to AB sliding pathway and $E_{\perp}$ can then be expressed as 
    \begin{equation}
        \label{eqn_Energy}
        \varepsilon(t) = \varepsilon_0(t) - E_{\parallel} \cdot p_{\parallel}(t) - E_{\perp} \cdot p_{\perp}(t).
    \end{equation}
    If we define the critical field as the minimum field at which the local energy minimum of $\varepsilon(t)$ disappears [i.e., $\mathrm{d}\varepsilon(t)/\mathrm{d}t \leq 0$, see the top inset of Fig.~\ref{fig:4-E_MD}(b)] along the BA to SP sliding pathway. Then, $E_{\perp, c}$ along $-z$ can be obtained analytically for a given $E_{\parallel}$ by solving Eq.~\ref{eqn_Energy} [see the purple line in Fig.~\ref{fig:4-E_MD}(b) and SM \cite{SM}]. When $E_{\parallel}$ is small, $E_{\perp, c}$ can be expressed in terms of $\sqrt E_{\parallel}$ as 
    \begin{equation}
    \begin{aligned}
        \label{eqn_Eperp}
        E_{\perp, c} =& - \frac{4 e_{4} t_{\mathrm{BA}}} {3 p_{3}^{\perp}} \left( 1-\sqrt{-\frac{p_{2}^{\parallel}+2 p_{4}^{\parallel} t_{\mathrm{BA}}^{2}} {e_{4} t_{\mathrm{BA}}^{2}}} \sqrt{E_{\parallel}} \right)\\
        =& E^0_{\perp, c}(1-\beta \sqrt{E_{\parallel}}),
    \end{aligned}
    \end{equation}
    where $\beta >0$. Clearly, Eq.~\ref{eqn_Eperp} shows that the presence of $E_{\parallel}$ will reduce the required $E_{\perp, c}$. In the absence of $E_{\parallel}$, $E^0_{\perp, c} = - 4 e_{4} t_{\mathrm{BA}}/ 3 p_{3}^{\perp} =$ 2.86 V/{\AA} is the vertical coercive field. More importantly, it is further found that the presence of $E_{\parallel}$ can lower the total coercive field $E_{t, c}$. When $E_{\parallel} \to 0$, $E_{t, c}$ can be expressed as 
    \begin{equation}
        \begin{aligned}
        E_{t, c} &= \sqrt{(E_{\perp, c})^2 + (E_{\parallel})^2}  \\
        &\to E^0_{\perp, c}\left(1-\beta \sqrt{E_{\parallel}}\right) < E^0_{\perp, c},
        \end{aligned}
    \end{equation}
    which clearly demonstrates that $E_{t, c}$ is smaller than $E^0_{\perp, c}$ with finite $E_{\parallel}$. Alternatively, if we define the critical field as the minimum field such that $\varepsilon(t)$ does not exceed that at BA [i.e., $\varepsilon(t) \leq \varepsilon(t_{\mathrm BA})$, see the bottom inset of Fig.~\ref{fig:4-E_MD}(b)], $E_{\perp, c}$ can also be solved numerically for a given $E_{\parallel}$, as shown by the green line in Fig.~\ref{fig:4-E_MD}(b). In both analytical and numerical cases, the minimum $E_{t, c}$ is reduced by more than 64\% and 70\%, respectively, comparing to the $E^0_{\perp, c}$.

    Then, we conduct MD simulations to verify the prediction that finite $E_{\parallel}$ can lower $E_{\perp, c}$ and $E_{t, c}$. By applying a constant $E_{\parallel}$ of 0.2 V/{\AA} along $\theta=60^{\circ}$ direction, $E_{\perp, c}$ is found to significantly reduce, comparing to $E_{\parallel}=0$ [see Fig.~\ref{fig:4-E_MD}(a)]. As in the previous analysis, this adjustment ensured that sliding occurred exclusively at $\theta = 60^{\circ}$. However, insufficient $E_{\parallel}$ magnitude may not adequately counteract random thermal fluctuations, potentially resulting in alternative sliding directions. 
    
    To determine the configurations of $E_{t, c}$ necessary to initiate sliding, we apply an inclined electric field to the $\mathrm{BN}$ bilayer at different temperatures. The field has varying in-plane and vertical components, adjusted in increments of 0.05 V/{\AA}, to initiate sliding within 30 ps [refer to Fig.S5]. At 0.1 K, the configuration of $E_{t, c}$ is shown by the yellow line in Fig.~\ref{fig:4-E_MD}(b), exhibiting a trend consistent with previous theoretical analyses. As shown in Fig.~\ref{fig:4-E_MD}(c), the minimum $E_{t, c}$ decreases by over 70\% as the temperature decreases, compared to the vertical $E_{t, c}$. It is noteworthy that experimentally measured coercive fields are often significantly lower than theoretically predicted values. This discrepancy, potentially related to the Landauer paradox \cite{Landauer1957, XuBin2017, XuBin2017_2}, may be due to the presence of inhomogeneities in the experimental samples \cite{Daumont2012, Lisenkov2009}. On another aspect, dip angle of the minimum $E_{t, c}$ at different temperatures almost keep constant of $24^{\circ}$. Such constant behavior will benefit to experimental setups and also to device design.

    {\it Discussion.} The decrease in $E_{t, c}$ with applying an inclined electric field relies on the presence of in-plane polarization along the sliding pathway. According to symmetry analysis \cite{JiJunyi2023}, bilayers, which are made of monolayers with the same layer group of $p \bar{6} m 2$ as $\mathrm{BN}$, all display the required in-plane polarization. Such discovery indicates that the strategy of inclined field can be generalized to different sliding systems, such as widely studied transition metal dichalcogenides (TMDs) $\mathrm{WSe_2}$, $\mathrm{MoSe_2}$, $\mathrm{WS_2}$, and $\mathrm{MoS_2}$ \cite{WangXirui2022, SunYan2022, BianRenji2024}, as well as $\mathrm{InSe}$ \cite{HuHaowen2019, SuiFengrui2023}. To confirm this prediction, we verify the energy and polarization distribution of $\mathrm{MoS_2}$ [refer to Fig.S6], which differs from $\mathrm{BN}$ only in the opposite direction of in-plane polarization. Hence, the direction of $E_{\parallel}$ need to be reversed to achieve efficient switching.

    In conclusion, our newly developed machine-learning-assisted DREAM-Allegro model effectively predicts the interatomic energy and BECs of parallel-stacked $\mathrm{BN}$ bilayers. This enables accurate prediction on temperature and field effects by performing MD simulations. Applying both simulations and theoretical analysis, we find that in-plane field can lead to deterministic switching of polarization and an inclined field can largely reduce the critical field required for switching. Such strategy is further generated to widely studied sliding ferroelectrics and thus paves the way for applications.

\begin{acknowledgments}
    We acknowledge financial support from the National Key R\&D Program of China (No. 2022YFA1402901), NSFC (No. 11825403, 11991061, 12188101, 12174060, and 12274082), the Guangdong Major Project of the Basic and Applied Basic Research (Future functional materials under extreme conditions—2021B0301030005), Shanghai Science and Technology Program (No. 23JC1400900), and Shanghai Pilot Program for Basic Research—FuDan University 21TQ1400100 (23TQ017). C.X. also acknowledges support from the Shanghai Science and Technology Committee (Grant No. 23ZR1406600) and the Xiaomi Young Talents Program.
\end{acknowledgments}

\bibliography{bilayerBN}

\begin{thebibliography}{60}%
\makeatletter
\providecommand \@ifxundefined [1]{%
 \@ifx{#1\undefined}
}%
\providecommand \@ifnum [1]{%
 \ifnum #1\expandafter \@firstoftwo
 \else \expandafter \@secondoftwo
 \fi
}%
\providecommand \@ifx [1]{%
 \ifx #1\expandafter \@firstoftwo
 \else \expandafter \@secondoftwo
 \fi
}%
\providecommand \natexlab [1]{#1}%
\providecommand \enquote  [1]{``#1''}%
\providecommand \bibnamefont  [1]{#1}%
\providecommand \bibfnamefont [1]{#1}%
\providecommand \citenamefont [1]{#1}%
\providecommand \href@noop [0]{\@secondoftwo}%
\providecommand \href [0]{\begingroup \@sanitize@url \@href}%
\providecommand \@href[1]{\@@startlink{#1}\@@href}%
\providecommand \@@href[1]{\endgroup#1\@@endlink}%
\providecommand \@sanitize@url [0]{\catcode `\\12\catcode `\$12\catcode `\&12\catcode `\#12\catcode `\^12\catcode `\_12\catcode `\%12\relax}%
\providecommand \@@startlink[1]{}%
\providecommand \@@endlink[0]{}%
\providecommand \url  [0]{\begingroup\@sanitize@url \@url }%
\providecommand \@url [1]{\endgroup\@href {#1}{\urlprefix }}%
\providecommand \urlprefix  [0]{URL }%
\providecommand \Eprint [0]{\href }%
\providecommand \doibase [0]{https://doi.org/}%
\providecommand \selectlanguage [0]{\@gobble}%
\providecommand \bibinfo  [0]{\@secondoftwo}%
\providecommand \bibfield  [0]{\@secondoftwo}%
\providecommand \translation [1]{[#1]}%
\providecommand \BibitemOpen [0]{}%
\providecommand \bibitemStop [0]{}%
\providecommand \bibitemNoStop [0]{.\EOS\space}%
\providecommand \EOS [0]{\spacefactor3000\relax}%
\providecommand \BibitemShut  [1]{\csname bibitem#1\endcsname}%
\let\auto@bib@innerbib\@empty
\bibitem [{\citenamefont {Yasuda}\ \emph {et~al.}(2021)\citenamefont {Yasuda}, \citenamefont {Wang}, \citenamefont {Watanabe}, \citenamefont {Taniguchi},\ and\ \citenamefont {Jarillo-Herrero}}]{Yasuda2021}%
  \BibitemOpen
  \bibfield  {author} {\bibinfo {author} {\bibfnamefont {K.}~\bibnamefont {Yasuda}}, \bibinfo {author} {\bibfnamefont {X.}~\bibnamefont {Wang}}, \bibinfo {author} {\bibfnamefont {K.}~\bibnamefont {Watanabe}}, \bibinfo {author} {\bibfnamefont {T.}~\bibnamefont {Taniguchi}},\ and\ \bibinfo {author} {\bibfnamefont {P.}~\bibnamefont {Jarillo-Herrero}},\ }\bibfield  {title} {\bibinfo {title} {Stacking-engineered ferroelectricity in bilayer boron nitride},\ }\href {https://doi.org/10.1126/science.abd3230} {\bibfield  {journal} {\bibinfo  {journal} {Science}\ }\textbf {\bibinfo {volume} {372}},\ \bibinfo {pages} {1458} (\bibinfo {year} {2021})},\ \Eprint {https://arxiv.org/abs/https://www.science.org/doi/pdf/10.1126/science.abd3230} {https://www.science.org/doi/pdf/10.1126/science.abd3230} \BibitemShut {NoStop}%
\bibitem [{\citenamefont {Stern}\ \emph {et~al.}(2021)\citenamefont {Stern}, \citenamefont {Waschitz}, \citenamefont {Cao}, \citenamefont {Nevo}, \citenamefont {Watanabe}, \citenamefont {Taniguchi}, \citenamefont {Sela}, \citenamefont {Urbakh}, \citenamefont {Hod},\ and\ \citenamefont {Shalom}}]{Stern2021}%
  \BibitemOpen
  \bibfield  {author} {\bibinfo {author} {\bibfnamefont {M.~V.}\ \bibnamefont {Stern}}, \bibinfo {author} {\bibfnamefont {Y.}~\bibnamefont {Waschitz}}, \bibinfo {author} {\bibfnamefont {W.}~\bibnamefont {Cao}}, \bibinfo {author} {\bibfnamefont {I.}~\bibnamefont {Nevo}}, \bibinfo {author} {\bibfnamefont {K.}~\bibnamefont {Watanabe}}, \bibinfo {author} {\bibfnamefont {T.}~\bibnamefont {Taniguchi}}, \bibinfo {author} {\bibfnamefont {E.}~\bibnamefont {Sela}}, \bibinfo {author} {\bibfnamefont {M.}~\bibnamefont {Urbakh}}, \bibinfo {author} {\bibfnamefont {O.}~\bibnamefont {Hod}},\ and\ \bibinfo {author} {\bibfnamefont {M.~B.}\ \bibnamefont {Shalom}},\ }\bibfield  {title} {\bibinfo {title} {Interfacial ferroelectricity by {van der Waals} sliding},\ }\href {https://doi.org/10.1126/science.abe8177} {\bibfield  {journal} {\bibinfo  {journal} {Science}\ }\textbf {\bibinfo {volume} {372}},\ \bibinfo {pages} {1462} (\bibinfo {year} {2021})},\ \Eprint {https://arxiv.org/abs/https://www.science.org/doi/pdf/10.1126/science.abe8177} {https://www.science.org/doi/pdf/10.1126/science.abe8177} \BibitemShut {NoStop}%
\bibitem [{\citenamefont {Müller}\ \emph {et~al.}(2015)\citenamefont {Müller}, \citenamefont {Polakowski}, \citenamefont {Mueller},\ and\ \citenamefont {Mikolajick}}]{Müller2015}%
  \BibitemOpen
  \bibfield  {author} {\bibinfo {author} {\bibfnamefont {J.}~\bibnamefont {Müller}}, \bibinfo {author} {\bibfnamefont {P.}~\bibnamefont {Polakowski}}, \bibinfo {author} {\bibfnamefont {S.}~\bibnamefont {Mueller}},\ and\ \bibinfo {author} {\bibfnamefont {T.}~\bibnamefont {Mikolajick}},\ }\bibfield  {title} {\bibinfo {title} {Ferroelectric hafnium oxide based materials and devices: Assessment of current status and future prospects},\ }\href {https://doi.org/10.1149/2.0081505jss} {\bibfield  {journal} {\bibinfo  {journal} {ECS Journal of Solid State Science and Technology}\ }\textbf {\bibinfo {volume} {4}},\ \bibinfo {pages} {N30} (\bibinfo {year} {2015})}\BibitemShut {NoStop}%
\bibitem [{\citenamefont {Wang}\ \emph {et~al.}(2023)\citenamefont {Wang}, \citenamefont {You}, \citenamefont {Cobden},\ and\ \citenamefont {Wang}}]{Wang2023}%
  \BibitemOpen
  \bibfield  {author} {\bibinfo {author} {\bibfnamefont {C.}~\bibnamefont {Wang}}, \bibinfo {author} {\bibfnamefont {L.}~\bibnamefont {You}}, \bibinfo {author} {\bibfnamefont {D.}~\bibnamefont {Cobden}},\ and\ \bibinfo {author} {\bibfnamefont {J.}~\bibnamefont {Wang}},\ }\bibfield  {title} {\bibinfo {title} {Towards two-dimensional {van der Waals} ferroelectrics},\ }\href {https://doi.org/10.1038/s41563-022-01422-y} {\bibfield  {journal} {\bibinfo  {journal} {Nature Materials}\ }\textbf {\bibinfo {volume} {22}},\ \bibinfo {pages} {542} (\bibinfo {year} {2023})}\BibitemShut {NoStop}%
\bibitem [{\citenamefont {Stengel}\ and\ \citenamefont {Spaldin}(2006)}]{Stengel2006}%
  \BibitemOpen
  \bibfield  {author} {\bibinfo {author} {\bibfnamefont {M.}~\bibnamefont {Stengel}}\ and\ \bibinfo {author} {\bibfnamefont {N.~A.}\ \bibnamefont {Spaldin}},\ }\bibfield  {title} {\bibinfo {title} {Origin of the dielectric dead layer in nanoscale capacitors},\ }\href {https://doi.org/10.1038/nature05148} {\bibfield  {journal} {\bibinfo  {journal} {Nature}\ }\textbf {\bibinfo {volume} {443}},\ \bibinfo {pages} {679} (\bibinfo {year} {2006})}\BibitemShut {NoStop}%
\bibitem [{\citenamefont {Li}\ and\ \citenamefont {Wu}(2017)}]{LiLei2017}%
  \BibitemOpen
  \bibfield  {author} {\bibinfo {author} {\bibfnamefont {L.}~\bibnamefont {Li}}\ and\ \bibinfo {author} {\bibfnamefont {M.}~\bibnamefont {Wu}},\ }\bibfield  {title} {\bibinfo {title} {Binary compound bilayer and multilayer with vertical polarizations: Two-dimensional ferroelectrics, multiferroics, and nanogenerators},\ }\href {https://doi.org/10.1021/acsnano.7b02756} {\bibfield  {journal} {\bibinfo  {journal} {ACS Nano}\ }\textbf {\bibinfo {volume} {11}},\ \bibinfo {pages} {6382} (\bibinfo {year} {2017})}\BibitemShut {NoStop}%
\bibitem [{\citenamefont {Ribeiro}\ and\ \citenamefont {Peres}(2011)}]{Ribeiro2011}%
  \BibitemOpen
  \bibfield  {author} {\bibinfo {author} {\bibfnamefont {R.~M.}\ \bibnamefont {Ribeiro}}\ and\ \bibinfo {author} {\bibfnamefont {N.~M.~R.}\ \bibnamefont {Peres}},\ }\bibfield  {title} {\bibinfo {title} {Stability of boron nitride bilayers: Ground-state energies, interlayer distances, and tight-binding description},\ }\href {https://doi.org/10.1103/PhysRevB.83.235312} {\bibfield  {journal} {\bibinfo  {journal} {Phys. Rev. B}\ }\textbf {\bibinfo {volume} {83}},\ \bibinfo {pages} {235312} (\bibinfo {year} {2011})}\BibitemShut {NoStop}%
\bibitem [{\citenamefont {Wu}\ and\ \citenamefont {Li}(2021)}]{WuMenghao2021_2}%
  \BibitemOpen
  \bibfield  {author} {\bibinfo {author} {\bibfnamefont {M.}~\bibnamefont {Wu}}\ and\ \bibinfo {author} {\bibfnamefont {J.}~\bibnamefont {Li}},\ }\bibfield  {title} {\bibinfo {title} {Sliding ferroelectricity in 2d {van der Waals} materials: Related physics and future opportunities},\ }\href {https://doi.org/10.1073/pnas.2115703118} {\bibfield  {journal} {\bibinfo  {journal} {Proceedings of the National Academy of Sciences}\ }\textbf {\bibinfo {volume} {118}},\ \bibinfo {pages} {e2115703118} (\bibinfo {year} {2021})},\ \Eprint {https://arxiv.org/abs/https://www.pnas.org/doi/pdf/10.1073/pnas.2115703118} {https://www.pnas.org/doi/pdf/10.1073/pnas.2115703118} \BibitemShut {NoStop}%
\bibitem [{\citenamefont {Constantinescu}\ \emph {et~al.}(2013)\citenamefont {Constantinescu}, \citenamefont {Kuc},\ and\ \citenamefont {Heine}}]{Constantinescu2013}%
  \BibitemOpen
  \bibfield  {author} {\bibinfo {author} {\bibfnamefont {G.}~\bibnamefont {Constantinescu}}, \bibinfo {author} {\bibfnamefont {A.}~\bibnamefont {Kuc}},\ and\ \bibinfo {author} {\bibfnamefont {T.}~\bibnamefont {Heine}},\ }\bibfield  {title} {\bibinfo {title} {Stacking in bulk and bilayer hexagonal boron nitride},\ }\href {https://doi.org/10.1103/PhysRevLett.111.036104} {\bibfield  {journal} {\bibinfo  {journal} {Phys. Rev. Lett.}\ }\textbf {\bibinfo {volume} {111}},\ \bibinfo {pages} {036104} (\bibinfo {year} {2013})}\BibitemShut {NoStop}%
\bibitem [{\citenamefont {Tsymbal}(2021)}]{Evgeny2021}%
  \BibitemOpen
  \bibfield  {author} {\bibinfo {author} {\bibfnamefont {E.~Y.}\ \bibnamefont {Tsymbal}},\ }\bibfield  {title} {\bibinfo {title} {Two-dimensional ferroelectricity by design},\ }\href {https://doi.org/10.1126/science.abi7296} {\bibfield  {journal} {\bibinfo  {journal} {Science}\ }\textbf {\bibinfo {volume} {372}},\ \bibinfo {pages} {1389} (\bibinfo {year} {2021})},\ \Eprint {https://arxiv.org/abs/https://www.science.org/doi/pdf/10.1126/science.abi7296} {https://www.science.org/doi/pdf/10.1126/science.abi7296} \BibitemShut {NoStop}%
\bibitem [{\citenamefont {Jiang}\ \emph {et~al.}(2022)\citenamefont {Jiang}, \citenamefont {Liu}, \citenamefont {Ma}, \citenamefont {Yu}, \citenamefont {Hu}, \citenamefont {Li}, \citenamefont {Burton}, \citenamefont {Liu}, \citenamefont {Chen}, \citenamefont {Guo}, \citenamefont {Kong}, \citenamefont {Bellaiche},\ and\ \citenamefont {Ren}}]{JiangWen2022}%
  \BibitemOpen
  \bibfield  {author} {\bibinfo {author} {\bibfnamefont {W.}~\bibnamefont {Jiang}}, \bibinfo {author} {\bibfnamefont {C.}~\bibnamefont {Liu}}, \bibinfo {author} {\bibfnamefont {X.}~\bibnamefont {Ma}}, \bibinfo {author} {\bibfnamefont {X.}~\bibnamefont {Yu}}, \bibinfo {author} {\bibfnamefont {S.}~\bibnamefont {Hu}}, \bibinfo {author} {\bibfnamefont {X.}~\bibnamefont {Li}}, \bibinfo {author} {\bibfnamefont {L.~A.}\ \bibnamefont {Burton}}, \bibinfo {author} {\bibfnamefont {Y.}~\bibnamefont {Liu}}, \bibinfo {author} {\bibfnamefont {Y.}~\bibnamefont {Chen}}, \bibinfo {author} {\bibfnamefont {P.}~\bibnamefont {Guo}}, \bibinfo {author} {\bibfnamefont {X.}~\bibnamefont {Kong}}, \bibinfo {author} {\bibfnamefont {L.}~\bibnamefont {Bellaiche}},\ and\ \bibinfo {author} {\bibfnamefont {W.}~\bibnamefont {Ren}},\ }\bibfield  {title} {\bibinfo {title} {Anomalous ferroelectricity and double-negative effects in bilayer hexagonal boron nitride},\ }\href {https://doi.org/10.1103/PhysRevB.106.054104} {\bibfield  {journal} {\bibinfo  {journal} {Phys. Rev. B}\ }\textbf {\bibinfo {volume} {106}},\ \bibinfo {pages} {054104} (\bibinfo {year} {2022})}\BibitemShut {NoStop}%
\bibitem [{\citenamefont {Zheng}\ \emph {et~al.}(2020)\citenamefont {Zheng}, \citenamefont {Ma}, \citenamefont {Bi}, \citenamefont {de~la Barrera}, \citenamefont {Liu}, \citenamefont {Mao}, \citenamefont {Zhang}, \citenamefont {Kiper}, \citenamefont {Watanabe}, \citenamefont {Taniguchi}, \citenamefont {Kong}, \citenamefont {Tisdale}, \citenamefont {Ashoori}, \citenamefont {Gedik}, \citenamefont {Fu}, \citenamefont {Xu},\ and\ \citenamefont {Jarillo-Herrero}}]{ZhengZhiren2020}%
  \BibitemOpen
  \bibfield  {author} {\bibinfo {author} {\bibfnamefont {Z.}~\bibnamefont {Zheng}}, \bibinfo {author} {\bibfnamefont {Q.}~\bibnamefont {Ma}}, \bibinfo {author} {\bibfnamefont {Z.}~\bibnamefont {Bi}}, \bibinfo {author} {\bibfnamefont {S.}~\bibnamefont {de~la Barrera}}, \bibinfo {author} {\bibfnamefont {M.-H.}\ \bibnamefont {Liu}}, \bibinfo {author} {\bibfnamefont {N.}~\bibnamefont {Mao}}, \bibinfo {author} {\bibfnamefont {Y.}~\bibnamefont {Zhang}}, \bibinfo {author} {\bibfnamefont {N.}~\bibnamefont {Kiper}}, \bibinfo {author} {\bibfnamefont {K.}~\bibnamefont {Watanabe}}, \bibinfo {author} {\bibfnamefont {T.}~\bibnamefont {Taniguchi}}, \bibinfo {author} {\bibfnamefont {J.}~\bibnamefont {Kong}}, \bibinfo {author} {\bibfnamefont {W.~A.}\ \bibnamefont {Tisdale}}, \bibinfo {author} {\bibfnamefont {R.}~\bibnamefont {Ashoori}}, \bibinfo {author} {\bibfnamefont {N.}~\bibnamefont {Gedik}}, \bibinfo {author} {\bibfnamefont {L.}~\bibnamefont {Fu}}, \bibinfo {author} {\bibfnamefont {S.-Y.}\ \bibnamefont {Xu}},\ and\ \bibinfo {author} {\bibfnamefont {P.}~\bibnamefont {Jarillo-Herrero}},\ }\bibfield  {title} {\bibinfo {title} {Unconventional ferroelectricity in moir{\'e} heterostructures},\ }\href {https://doi.org/10.1038/s41586-020-2970-9} {\bibfield  {journal} {\bibinfo  {journal} {Nature}\ }\textbf {\bibinfo {volume} {588}},\ \bibinfo {pages} {71} (\bibinfo {year} {2020})}\BibitemShut {NoStop}%
\bibitem [{\citenamefont {Liu}\ \emph {et~al.}(2023)\citenamefont {Liu}, \citenamefont {Ji}, \citenamefont {Fu}, \citenamefont {Wang}, \citenamefont {Sun},\ and\ \citenamefont {Gao}}]{LiuMeng2023}%
  \BibitemOpen
  \bibfield  {author} {\bibinfo {author} {\bibfnamefont {M.}~\bibnamefont {Liu}}, \bibinfo {author} {\bibfnamefont {H.}~\bibnamefont {Ji}}, \bibinfo {author} {\bibfnamefont {Z.}~\bibnamefont {Fu}}, \bibinfo {author} {\bibfnamefont {Y.}~\bibnamefont {Wang}}, \bibinfo {author} {\bibfnamefont {J.-T.}\ \bibnamefont {Sun}},\ and\ \bibinfo {author} {\bibfnamefont {H.-J.}\ \bibnamefont {Gao}},\ }\bibfield  {title} {\bibinfo {title} {Orbital distortion and electric field control of sliding ferroelectricity in a boron nitride bilayer},\ }\href {https://doi.org/10.1088/1361-648X/acc561} {\bibfield  {journal} {\bibinfo  {journal} {Journal of Physics: Condensed Matter}\ }\textbf {\bibinfo {volume} {35}},\ \bibinfo {pages} {235001} (\bibinfo {year} {2023})}\BibitemShut {NoStop}%
\bibitem [{\citenamefont {Tang}\ and\ \citenamefont {Bauer}(2023)}]{TangPing2023}%
  \BibitemOpen
  \bibfield  {author} {\bibinfo {author} {\bibfnamefont {P.}~\bibnamefont {Tang}}\ and\ \bibinfo {author} {\bibfnamefont {G.~E.~W.}\ \bibnamefont {Bauer}},\ }\bibfield  {title} {\bibinfo {title} {Sliding phase transition in ferroelectric {van der Waals} bilayers},\ }\href {https://doi.org/10.1103/PhysRevLett.130.176801} {\bibfield  {journal} {\bibinfo  {journal} {Phys. Rev. Lett.}\ }\textbf {\bibinfo {volume} {130}},\ \bibinfo {pages} {176801} (\bibinfo {year} {2023})}\BibitemShut {NoStop}%
\bibitem [{\citenamefont {Ji}\ \emph {et~al.}(2023)\citenamefont {Ji}, \citenamefont {Yu}, \citenamefont {Xu},\ and\ \citenamefont {Xiang}}]{JiJunyi2023}%
  \BibitemOpen
  \bibfield  {author} {\bibinfo {author} {\bibfnamefont {J.}~\bibnamefont {Ji}}, \bibinfo {author} {\bibfnamefont {G.}~\bibnamefont {Yu}}, \bibinfo {author} {\bibfnamefont {C.}~\bibnamefont {Xu}},\ and\ \bibinfo {author} {\bibfnamefont {H.~J.}\ \bibnamefont {Xiang}},\ }\bibfield  {title} {\bibinfo {title} {General theory for bilayer stacking ferroelectricity},\ }\href {https://doi.org/10.1103/PhysRevLett.130.146801} {\bibfield  {journal} {\bibinfo  {journal} {Phys. Rev. Lett.}\ }\textbf {\bibinfo {volume} {130}},\ \bibinfo {pages} {146801} (\bibinfo {year} {2023})}\BibitemShut {NoStop}%
\bibitem [{\citenamefont {Fei}\ \emph {et~al.}(2018)\citenamefont {Fei}, \citenamefont {Zhao}, \citenamefont {Palomaki}, \citenamefont {Sun}, \citenamefont {Miller}, \citenamefont {Zhao}, \citenamefont {Yan}, \citenamefont {Xu},\ and\ \citenamefont {Cobden}}]{FeiZaiyao2018}%
  \BibitemOpen
  \bibfield  {author} {\bibinfo {author} {\bibfnamefont {Z.}~\bibnamefont {Fei}}, \bibinfo {author} {\bibfnamefont {W.}~\bibnamefont {Zhao}}, \bibinfo {author} {\bibfnamefont {T.~A.}\ \bibnamefont {Palomaki}}, \bibinfo {author} {\bibfnamefont {B.}~\bibnamefont {Sun}}, \bibinfo {author} {\bibfnamefont {M.~K.}\ \bibnamefont {Miller}}, \bibinfo {author} {\bibfnamefont {Z.}~\bibnamefont {Zhao}}, \bibinfo {author} {\bibfnamefont {J.}~\bibnamefont {Yan}}, \bibinfo {author} {\bibfnamefont {X.}~\bibnamefont {Xu}},\ and\ \bibinfo {author} {\bibfnamefont {D.~H.}\ \bibnamefont {Cobden}},\ }\bibfield  {title} {\bibinfo {title} {Ferroelectric switching of a two-dimensional metal},\ }\href {https://doi.org/10.1038/s41586-018-0336-3} {\bibfield  {journal} {\bibinfo  {journal} {Nature}\ }\textbf {\bibinfo {volume} {560}},\ \bibinfo {pages} {336} (\bibinfo {year} {2018})}\BibitemShut {NoStop}%
\bibitem [{\citenamefont {Sharma}\ \emph {et~al.}(2019)\citenamefont {Sharma}, \citenamefont {Xiang}, \citenamefont {Shao}, \citenamefont {Zhang}, \citenamefont {Tsymbal}, \citenamefont {Hamilton},\ and\ \citenamefont {Seidel}}]{Sharma2019}%
  \BibitemOpen
  \bibfield  {author} {\bibinfo {author} {\bibfnamefont {P.}~\bibnamefont {Sharma}}, \bibinfo {author} {\bibfnamefont {F.-X.}\ \bibnamefont {Xiang}}, \bibinfo {author} {\bibfnamefont {D.-F.}\ \bibnamefont {Shao}}, \bibinfo {author} {\bibfnamefont {D.}~\bibnamefont {Zhang}}, \bibinfo {author} {\bibfnamefont {E.~Y.}\ \bibnamefont {Tsymbal}}, \bibinfo {author} {\bibfnamefont {A.~R.}\ \bibnamefont {Hamilton}},\ and\ \bibinfo {author} {\bibfnamefont {J.}~\bibnamefont {Seidel}},\ }\bibfield  {title} {\bibinfo {title} {A room-temperature ferroelectric semimetal},\ }\href {https://doi.org/10.1126/sciadv.aax5080} {\bibfield  {journal} {\bibinfo  {journal} {Science advances}\ }\textbf {\bibinfo {volume} {5}},\ \bibinfo {pages} {eaax5080} (\bibinfo {year} {2019})}\BibitemShut {NoStop}%
\bibitem [{\citenamefont {Xiao}\ \emph {et~al.}(2020)\citenamefont {Xiao}, \citenamefont {Wang}, \citenamefont {Wang}, \citenamefont {Pemmaraju}, \citenamefont {Wang}, \citenamefont {Muscher}, \citenamefont {Sie}, \citenamefont {Nyby}, \citenamefont {Devereaux}, \citenamefont {Qian}, \citenamefont {Zhang},\ and\ \citenamefont {Lindenberg}}]{XiaoJun2020}%
  \BibitemOpen
  \bibfield  {author} {\bibinfo {author} {\bibfnamefont {J.}~\bibnamefont {Xiao}}, \bibinfo {author} {\bibfnamefont {Y.}~\bibnamefont {Wang}}, \bibinfo {author} {\bibfnamefont {H.}~\bibnamefont {Wang}}, \bibinfo {author} {\bibfnamefont {C.~D.}\ \bibnamefont {Pemmaraju}}, \bibinfo {author} {\bibfnamefont {S.}~\bibnamefont {Wang}}, \bibinfo {author} {\bibfnamefont {P.}~\bibnamefont {Muscher}}, \bibinfo {author} {\bibfnamefont {E.~J.}\ \bibnamefont {Sie}}, \bibinfo {author} {\bibfnamefont {C.~M.}\ \bibnamefont {Nyby}}, \bibinfo {author} {\bibfnamefont {T.~P.}\ \bibnamefont {Devereaux}}, \bibinfo {author} {\bibfnamefont {X.}~\bibnamefont {Qian}}, \bibinfo {author} {\bibfnamefont {X.}~\bibnamefont {Zhang}},\ and\ \bibinfo {author} {\bibfnamefont {A.~M.}\ \bibnamefont {Lindenberg}},\ }\bibfield  {title} {\bibinfo {title} {Berry curvature memory through electrically driven stacking transitions},\ }\href {https://doi.org/10.1038/s41567-020-0947-0} {\bibfield  {journal} {\bibinfo  {journal} {Nature Physics}\ }\textbf {\bibinfo {volume} {16}},\ \bibinfo {pages} {1028} (\bibinfo {year} {2020})}\BibitemShut {NoStop}%
\bibitem [{\citenamefont {de~la Barrera}\ \emph {et~al.}(2021)\citenamefont {de~la Barrera}, \citenamefont {Cao}, \citenamefont {Gao}, \citenamefont {Gao}, \citenamefont {Bheemarasetty}, \citenamefont {Yan}, \citenamefont {Mandrus}, \citenamefont {Zhu}, \citenamefont {Xiao},\ and\ \citenamefont {Hunt}}]{Barrera2021}%
  \BibitemOpen
  \bibfield  {author} {\bibinfo {author} {\bibfnamefont {S.~C.}\ \bibnamefont {de~la Barrera}}, \bibinfo {author} {\bibfnamefont {Q.}~\bibnamefont {Cao}}, \bibinfo {author} {\bibfnamefont {Y.}~\bibnamefont {Gao}}, \bibinfo {author} {\bibfnamefont {Y.}~\bibnamefont {Gao}}, \bibinfo {author} {\bibfnamefont {V.~S.}\ \bibnamefont {Bheemarasetty}}, \bibinfo {author} {\bibfnamefont {J.}~\bibnamefont {Yan}}, \bibinfo {author} {\bibfnamefont {D.~G.}\ \bibnamefont {Mandrus}}, \bibinfo {author} {\bibfnamefont {W.}~\bibnamefont {Zhu}}, \bibinfo {author} {\bibfnamefont {D.}~\bibnamefont {Xiao}},\ and\ \bibinfo {author} {\bibfnamefont {B.~M.}\ \bibnamefont {Hunt}},\ }\bibfield  {title} {\bibinfo {title} {Direct measurement of ferroelectric polarization in a tunable semimetal},\ }\href {https://doi.org/10.1038/s41467-021-25587-3} {\bibfield  {journal} {\bibinfo  {journal} {Nature Communications}\ }\textbf {\bibinfo {volume} {12}},\ \bibinfo {pages} {5298} (\bibinfo {year} {2021})}\BibitemShut {NoStop}%
\bibitem [{\citenamefont {Jindal}\ \emph {et~al.}(2023)\citenamefont {Jindal}, \citenamefont {Saha}, \citenamefont {Li}, \citenamefont {Taniguchi}, \citenamefont {Watanabe}, \citenamefont {Hone}, \citenamefont {Birol}, \citenamefont {Fernandes}, \citenamefont {Dean}, \citenamefont {Pasupathy},\ and\ \citenamefont {Rhodes}}]{Jindal2023}%
  \BibitemOpen
  \bibfield  {author} {\bibinfo {author} {\bibfnamefont {A.}~\bibnamefont {Jindal}}, \bibinfo {author} {\bibfnamefont {A.}~\bibnamefont {Saha}}, \bibinfo {author} {\bibfnamefont {Z.}~\bibnamefont {Li}}, \bibinfo {author} {\bibfnamefont {T.}~\bibnamefont {Taniguchi}}, \bibinfo {author} {\bibfnamefont {K.}~\bibnamefont {Watanabe}}, \bibinfo {author} {\bibfnamefont {J.~C.}\ \bibnamefont {Hone}}, \bibinfo {author} {\bibfnamefont {T.}~\bibnamefont {Birol}}, \bibinfo {author} {\bibfnamefont {R.~M.}\ \bibnamefont {Fernandes}}, \bibinfo {author} {\bibfnamefont {C.~R.}\ \bibnamefont {Dean}}, \bibinfo {author} {\bibfnamefont {A.~N.}\ \bibnamefont {Pasupathy}},\ and\ \bibinfo {author} {\bibfnamefont {D.~A.}\ \bibnamefont {Rhodes}},\ }\bibfield  {title} {\bibinfo {title} {Coupled ferroelectricity and superconductivity in bilayer {Td-MoTe2}},\ }\href {https://doi.org/10.1038/s41586-022-05521-3} {\bibfield  {journal} {\bibinfo  {journal} {Nature}\ }\textbf {\bibinfo {volume} {613}},\ \bibinfo {pages} {48} (\bibinfo {year} {2023})}\BibitemShut {NoStop}%
\bibitem [{\citenamefont {Lv}\ \emph {et~al.}(2022)\citenamefont {Lv}, \citenamefont {Sun}, \citenamefont {Chen}, \citenamefont {Taniguchi}, \citenamefont {Watanabe}, \citenamefont {Wu}, \citenamefont {Wang},\ and\ \citenamefont {Xue}}]{LvMing2022}%
  \BibitemOpen
  \bibfield  {author} {\bibinfo {author} {\bibfnamefont {M.}~\bibnamefont {Lv}}, \bibinfo {author} {\bibfnamefont {X.}~\bibnamefont {Sun}}, \bibinfo {author} {\bibfnamefont {Y.}~\bibnamefont {Chen}}, \bibinfo {author} {\bibfnamefont {T.}~\bibnamefont {Taniguchi}}, \bibinfo {author} {\bibfnamefont {K.}~\bibnamefont {Watanabe}}, \bibinfo {author} {\bibfnamefont {M.}~\bibnamefont {Wu}}, \bibinfo {author} {\bibfnamefont {J.}~\bibnamefont {Wang}},\ and\ \bibinfo {author} {\bibfnamefont {J.}~\bibnamefont {Xue}},\ }\bibfield  {title} {\bibinfo {title} {Spatially resolved polarization manipulation of ferroelectricity in twisted {hBN}},\ }\href {https://doi.org/https://doi.org/10.1002/adma.202203990} {\bibfield  {journal} {\bibinfo  {journal} {Advanced Materials}\ }\textbf {\bibinfo {volume} {34}},\ \bibinfo {pages} {2203990} (\bibinfo {year} {2022})},\ \Eprint {https://arxiv.org/abs/https://onlinelibrary.wiley.com/doi/pdf/10.1002/adma.202203990} {https://onlinelibrary.wiley.com/doi/pdf/10.1002/adma.202203990} \BibitemShut {NoStop}%
\bibitem [{\citenamefont {Hu}\ \emph {et~al.}(2019)\citenamefont {Hu}, \citenamefont {Sun}, \citenamefont {Chai}, \citenamefont {Xie}, \citenamefont {Ma},\ and\ \citenamefont {Zhu}}]{HuHaowen2019}%
  \BibitemOpen
  \bibfield  {author} {\bibinfo {author} {\bibfnamefont {H.}~\bibnamefont {Hu}}, \bibinfo {author} {\bibfnamefont {Y.}~\bibnamefont {Sun}}, \bibinfo {author} {\bibfnamefont {M.}~\bibnamefont {Chai}}, \bibinfo {author} {\bibfnamefont {D.}~\bibnamefont {Xie}}, \bibinfo {author} {\bibfnamefont {J.}~\bibnamefont {Ma}},\ and\ \bibinfo {author} {\bibfnamefont {H.}~\bibnamefont {Zhu}},\ }\bibfield  {title} {\bibinfo {title} {{Room-temperature out-of-plane and in-plane ferroelectricity of two-dimensional $\beta$-InSe nanoflakes}},\ }\href {https://doi.org/10.1063/1.5097842} {\bibfield  {journal} {\bibinfo  {journal} {Applied Physics Letters}\ }\textbf {\bibinfo {volume} {114}},\ \bibinfo {pages} {252903} (\bibinfo {year} {2019})},\ \Eprint {https://arxiv.org/abs/https://pubs.aip.org/aip/apl/article-pdf/doi/10.1063/1.5097842/14525668/252903\_1\_online.pdf} {https://pubs.aip.org/aip/apl/article-pdf/doi/10.1063/1.5097842/14525668/252903\_1\_online.pdf} \BibitemShut {NoStop}%
\bibitem [{\citenamefont {Sui}\ \emph {et~al.}(2023)\citenamefont {Sui}, \citenamefont {Jin}, \citenamefont {Zhang}, \citenamefont {Qi}, \citenamefont {Wu}, \citenamefont {Huang}, \citenamefont {Yue},\ and\ \citenamefont {Chu}}]{SuiFengrui2023}%
  \BibitemOpen
  \bibfield  {author} {\bibinfo {author} {\bibfnamefont {F.}~\bibnamefont {Sui}}, \bibinfo {author} {\bibfnamefont {M.}~\bibnamefont {Jin}}, \bibinfo {author} {\bibfnamefont {Y.}~\bibnamefont {Zhang}}, \bibinfo {author} {\bibfnamefont {R.}~\bibnamefont {Qi}}, \bibinfo {author} {\bibfnamefont {Y.-N.}\ \bibnamefont {Wu}}, \bibinfo {author} {\bibfnamefont {R.}~\bibnamefont {Huang}}, \bibinfo {author} {\bibfnamefont {F.}~\bibnamefont {Yue}},\ and\ \bibinfo {author} {\bibfnamefont {J.}~\bibnamefont {Chu}},\ }\bibfield  {title} {\bibinfo {title} {Sliding ferroelectricity in {van der Waals} layered $\gamma$-{InSe} semiconductor},\ }\href {https://doi.org/10.1038/s41467-022-35490-0} {\bibfield  {journal} {\bibinfo  {journal} {Nature Communications}\ }\textbf {\bibinfo {volume} {14}},\ \bibinfo {pages} {36} (\bibinfo {year} {2023})}\BibitemShut {NoStop}%
\bibitem [{\citenamefont {Wang}\ \emph {et~al.}(2022)\citenamefont {Wang}, \citenamefont {Yasuda}, \citenamefont {Zhang}, \citenamefont {Liu}, \citenamefont {Watanabe}, \citenamefont {Taniguchi}, \citenamefont {Hone}, \citenamefont {Fu},\ and\ \citenamefont {Jarillo-Herrero}}]{WangXirui2022}%
  \BibitemOpen
  \bibfield  {author} {\bibinfo {author} {\bibfnamefont {X.}~\bibnamefont {Wang}}, \bibinfo {author} {\bibfnamefont {K.}~\bibnamefont {Yasuda}}, \bibinfo {author} {\bibfnamefont {Y.}~\bibnamefont {Zhang}}, \bibinfo {author} {\bibfnamefont {S.}~\bibnamefont {Liu}}, \bibinfo {author} {\bibfnamefont {K.}~\bibnamefont {Watanabe}}, \bibinfo {author} {\bibfnamefont {T.}~\bibnamefont {Taniguchi}}, \bibinfo {author} {\bibfnamefont {J.}~\bibnamefont {Hone}}, \bibinfo {author} {\bibfnamefont {L.}~\bibnamefont {Fu}},\ and\ \bibinfo {author} {\bibfnamefont {P.}~\bibnamefont {Jarillo-Herrero}},\ }\bibfield  {title} {\bibinfo {title} {Interfacial ferroelectricity in rhombohedral-stacked bilayer transition metal dichalcogenides},\ }\href {https://doi.org/10.1038/s41565-021-01059-z} {\bibfield  {journal} {\bibinfo  {journal} {Nature Nanotechnology}\ }\textbf {\bibinfo {volume} {17}},\ \bibinfo {pages} {367} (\bibinfo {year} {2022})}\BibitemShut {NoStop}%
\bibitem [{\citenamefont {Sun}\ \emph {et~al.}(2022)\citenamefont {Sun}, \citenamefont {Xu}, \citenamefont {Xu}, \citenamefont {Tian}, \citenamefont {Bai}, \citenamefont {Qi}, \citenamefont {Niu}, \citenamefont {Aung}, \citenamefont {Xiong}, \citenamefont {Han}, \citenamefont {Lu}, \citenamefont {Yin}, \citenamefont {Wang}, \citenamefont {Chen}, \citenamefont {Tenne}, \citenamefont {Zak},\ and\ \citenamefont {Guo}}]{SunYan2022}%
  \BibitemOpen
  \bibfield  {author} {\bibinfo {author} {\bibfnamefont {Y.}~\bibnamefont {Sun}}, \bibinfo {author} {\bibfnamefont {S.}~\bibnamefont {Xu}}, \bibinfo {author} {\bibfnamefont {Z.}~\bibnamefont {Xu}}, \bibinfo {author} {\bibfnamefont {J.}~\bibnamefont {Tian}}, \bibinfo {author} {\bibfnamefont {M.}~\bibnamefont {Bai}}, \bibinfo {author} {\bibfnamefont {Z.}~\bibnamefont {Qi}}, \bibinfo {author} {\bibfnamefont {Y.}~\bibnamefont {Niu}}, \bibinfo {author} {\bibfnamefont {H.~H.}\ \bibnamefont {Aung}}, \bibinfo {author} {\bibfnamefont {X.}~\bibnamefont {Xiong}}, \bibinfo {author} {\bibfnamefont {J.}~\bibnamefont {Han}}, \bibinfo {author} {\bibfnamefont {C.}~\bibnamefont {Lu}}, \bibinfo {author} {\bibfnamefont {J.}~\bibnamefont {Yin}}, \bibinfo {author} {\bibfnamefont {S.}~\bibnamefont {Wang}}, \bibinfo {author} {\bibfnamefont {Q.}~\bibnamefont {Chen}}, \bibinfo {author} {\bibfnamefont {R.}~\bibnamefont {Tenne}}, \bibinfo {author} {\bibfnamefont {A.}~\bibnamefont {Zak}},\ and\ \bibinfo {author} {\bibfnamefont {Y.}~\bibnamefont {Guo}},\ }\bibfield  {title} {\bibinfo {title} {Mesoscopic sliding ferroelectricity enabled photovoltaic random access memory for material-level artificial vision system},\ }\href {https://doi.org/10.1038/s41467-022-33118-x} {\bibfield  {journal} {\bibinfo  {journal} {Nature Communications}\ }\textbf {\bibinfo {volume} {13}},\ \bibinfo {pages} {5391} (\bibinfo {year} {2022})}\BibitemShut {NoStop}%
\bibitem [{\citenamefont {Wan}\ \emph {et~al.}(2022)\citenamefont {Wan}, \citenamefont {Hu}, \citenamefont {Mao}, \citenamefont {Fu}, \citenamefont {Yuan}, \citenamefont {Song}, \citenamefont {Gan}, \citenamefont {Xu}, \citenamefont {Xue}, \citenamefont {Cheng}, \citenamefont {Huang}, \citenamefont {Yang}, \citenamefont {Dai}, \citenamefont {Zeng},\ and\ \citenamefont {Kan}}]{WanYi2022}%
  \BibitemOpen
  \bibfield  {author} {\bibinfo {author} {\bibfnamefont {Y.}~\bibnamefont {Wan}}, \bibinfo {author} {\bibfnamefont {T.}~\bibnamefont {Hu}}, \bibinfo {author} {\bibfnamefont {X.}~\bibnamefont {Mao}}, \bibinfo {author} {\bibfnamefont {J.}~\bibnamefont {Fu}}, \bibinfo {author} {\bibfnamefont {K.}~\bibnamefont {Yuan}}, \bibinfo {author} {\bibfnamefont {Y.}~\bibnamefont {Song}}, \bibinfo {author} {\bibfnamefont {X.}~\bibnamefont {Gan}}, \bibinfo {author} {\bibfnamefont {X.}~\bibnamefont {Xu}}, \bibinfo {author} {\bibfnamefont {M.}~\bibnamefont {Xue}}, \bibinfo {author} {\bibfnamefont {X.}~\bibnamefont {Cheng}}, \bibinfo {author} {\bibfnamefont {C.}~\bibnamefont {Huang}}, \bibinfo {author} {\bibfnamefont {J.}~\bibnamefont {Yang}}, \bibinfo {author} {\bibfnamefont {L.}~\bibnamefont {Dai}}, \bibinfo {author} {\bibfnamefont {H.}~\bibnamefont {Zeng}},\ and\ \bibinfo {author} {\bibfnamefont {E.}~\bibnamefont {Kan}},\ }\bibfield  {title} {\bibinfo {title} {Room-temperature ferroelectricity in {$1{\mathrm{T}}^{\ensuremath{'}}$-${\mathrm{ReS}}_{2}$} multilayers},\ }\href {https://doi.org/10.1103/PhysRevLett.128.067601} {\bibfield  {journal} {\bibinfo  {journal} {Phys. Rev. Lett.}\ }\textbf {\bibinfo {volume} {128}},\ \bibinfo {pages} {067601} (\bibinfo {year} {2022})}\BibitemShut {NoStop}%
\bibitem [{\citenamefont {Miao}\ \emph {et~al.}(2022)\citenamefont {Miao}, \citenamefont {Ding}, \citenamefont {Wang}, \citenamefont {Shi}, \citenamefont {Ye}, \citenamefont {Li}, \citenamefont {Yao}, \citenamefont {Dong},\ and\ \citenamefont {Zhang}}]{MiaoLeping2022}%
  \BibitemOpen
  \bibfield  {author} {\bibinfo {author} {\bibfnamefont {L.-P.}\ \bibnamefont {Miao}}, \bibinfo {author} {\bibfnamefont {N.}~\bibnamefont {Ding}}, \bibinfo {author} {\bibfnamefont {N.}~\bibnamefont {Wang}}, \bibinfo {author} {\bibfnamefont {C.}~\bibnamefont {Shi}}, \bibinfo {author} {\bibfnamefont {H.-Y.}\ \bibnamefont {Ye}}, \bibinfo {author} {\bibfnamefont {L.}~\bibnamefont {Li}}, \bibinfo {author} {\bibfnamefont {Y.-F.}\ \bibnamefont {Yao}}, \bibinfo {author} {\bibfnamefont {S.}~\bibnamefont {Dong}},\ and\ \bibinfo {author} {\bibfnamefont {Y.}~\bibnamefont {Zhang}},\ }\bibfield  {title} {\bibinfo {title} {Direct observation of geometric and sliding ferroelectricity in an amphidynamic crystal},\ }\href {https://doi.org/10.1038/s41563-022-01322-1} {\bibfield  {journal} {\bibinfo  {journal} {Nature Materials}\ }\textbf {\bibinfo {volume} {21}},\ \bibinfo {pages} {1158} (\bibinfo {year} {2022})}\BibitemShut {NoStop}%
\bibitem [{\citenamefont {Yasuda}\ \emph {et~al.}(2024)\citenamefont {Yasuda}, \citenamefont {Zalys-Geller}, \citenamefont {Wang}, \citenamefont {Bennett}, \citenamefont {Cheema}, \citenamefont {Watanabe}, \citenamefont {Taniguchi}, \citenamefont {Kaxiras}, \citenamefont {Jarillo-Herrero},\ and\ \citenamefont {Ashoori}}]{Yasuda2024}%
  \BibitemOpen
  \bibfield  {author} {\bibinfo {author} {\bibfnamefont {K.}~\bibnamefont {Yasuda}}, \bibinfo {author} {\bibfnamefont {E.}~\bibnamefont {Zalys-Geller}}, \bibinfo {author} {\bibfnamefont {X.}~\bibnamefont {Wang}}, \bibinfo {author} {\bibfnamefont {D.}~\bibnamefont {Bennett}}, \bibinfo {author} {\bibfnamefont {S.~S.}\ \bibnamefont {Cheema}}, \bibinfo {author} {\bibfnamefont {K.}~\bibnamefont {Watanabe}}, \bibinfo {author} {\bibfnamefont {T.}~\bibnamefont {Taniguchi}}, \bibinfo {author} {\bibfnamefont {E.}~\bibnamefont {Kaxiras}}, \bibinfo {author} {\bibfnamefont {P.}~\bibnamefont {Jarillo-Herrero}},\ and\ \bibinfo {author} {\bibfnamefont {R.}~\bibnamefont {Ashoori}},\ }\bibfield  {title} {\bibinfo {title} {Ultrafast high-endurance memory based on sliding ferroelectrics},\ }\href {https://doi.org/10.1126/science.adp3575} {\bibfield  {journal} {\bibinfo  {journal} {Science}\ }\textbf {\bibinfo {volume} {0}},\ \bibinfo {pages} {eadp3575} (\bibinfo {year} {2024})}\BibitemShut {NoStop}%
\bibitem [{\citenamefont {Bian}\ \emph {et~al.}(2024)\citenamefont {Bian}, \citenamefont {He}, \citenamefont {Pan}, \citenamefont {Li}, \citenamefont {Cao}, \citenamefont {Meng}, \citenamefont {Chen}, \citenamefont {Liu}, \citenamefont {Zhong}, \citenamefont {Li},\ and\ \citenamefont {Liu}}]{BianRenji2024}%
  \BibitemOpen
  \bibfield  {author} {\bibinfo {author} {\bibfnamefont {R.}~\bibnamefont {Bian}}, \bibinfo {author} {\bibfnamefont {R.}~\bibnamefont {He}}, \bibinfo {author} {\bibfnamefont {E.}~\bibnamefont {Pan}}, \bibinfo {author} {\bibfnamefont {Z.}~\bibnamefont {Li}}, \bibinfo {author} {\bibfnamefont {G.}~\bibnamefont {Cao}}, \bibinfo {author} {\bibfnamefont {P.}~\bibnamefont {Meng}}, \bibinfo {author} {\bibfnamefont {J.}~\bibnamefont {Chen}}, \bibinfo {author} {\bibfnamefont {Q.}~\bibnamefont {Liu}}, \bibinfo {author} {\bibfnamefont {Z.}~\bibnamefont {Zhong}}, \bibinfo {author} {\bibfnamefont {W.}~\bibnamefont {Li}},\ and\ \bibinfo {author} {\bibfnamefont {F.}~\bibnamefont {Liu}},\ }\bibfield  {title} {\bibinfo {title} {Developing fatigue-resistant ferroelectrics using interlayer sliding switching},\ }\href {https://doi.org/10.1126/science.ado1744} {\bibfield  {journal} {\bibinfo  {journal} {Science}\ }\textbf {\bibinfo {volume} {0}},\ \bibinfo {pages} {eado1744} (\bibinfo {year} {2024})},\ \Eprint {https://arxiv.org/abs/https://www.science.org/doi/pdf/10.1126/science.ado1744} {https://www.science.org/doi/pdf/10.1126/science.ado1744} \BibitemShut {NoStop}%
\bibitem [{\citenamefont {Yang}\ \emph {et~al.}(2018)\citenamefont {Yang}, \citenamefont {Wu},\ and\ \citenamefont {Li}}]{YangQing2018}%
  \BibitemOpen
  \bibfield  {author} {\bibinfo {author} {\bibfnamefont {Q.}~\bibnamefont {Yang}}, \bibinfo {author} {\bibfnamefont {M.}~\bibnamefont {Wu}},\ and\ \bibinfo {author} {\bibfnamefont {J.}~\bibnamefont {Li}},\ }\bibfield  {title} {\bibinfo {title} {Origin of two-dimensional vertical ferroelectricity in {WTe2} bilayer and multilayer},\ }\href {https://doi.org/10.1021/acs.jpclett.8b03654} {\bibfield  {journal} {\bibinfo  {journal} {The Journal of Physical Chemistry Letters}\ }\textbf {\bibinfo {volume} {9}},\ \bibinfo {pages} {7160} (\bibinfo {year} {2018})}\BibitemShut {NoStop}%
\bibitem [{\citenamefont {Wu}(2021)}]{WuMenghao2021}%
  \BibitemOpen
  \bibfield  {author} {\bibinfo {author} {\bibfnamefont {M.}~\bibnamefont {Wu}},\ }\bibfield  {title} {\bibinfo {title} {Two-dimensional {van der Waals} ferroelectrics: Scientific and technological opportunities},\ }\href {https://doi.org/10.1021/acsnano.0c08483} {\bibfield  {journal} {\bibinfo  {journal} {ACS Nano}\ }\textbf {\bibinfo {volume} {15}},\ \bibinfo {pages} {9229} (\bibinfo {year} {2021})}\BibitemShut {NoStop}%
\bibitem [{\citenamefont {Bennett}\ and\ \citenamefont {Ghosez}(2024)}]{Bennett2024}%
  \BibitemOpen
  \bibfield  {author} {\bibinfo {author} {\bibfnamefont {D.}~\bibnamefont {Bennett}}\ and\ \bibinfo {author} {\bibfnamefont {P.}~\bibnamefont {Ghosez}},\ }\href {https://arxiv.org/abs/2404.10549} {\bibinfo {title} {Asymmetric dynamical charges in two-dimensional ferroelectrics}} (\bibinfo {year} {2024}),\ \Eprint {https://arxiv.org/abs/2404.10549} {arXiv:2404.10549 [cond-mat.mtrl-sci]} \BibitemShut {NoStop}%
\bibitem [{\citenamefont {He}\ \emph {et~al.}(2024)\citenamefont {He}, \citenamefont {Zhang}, \citenamefont {Wang}, \citenamefont {Li}, \citenamefont {Tang}, \citenamefont {Bauer},\ and\ \citenamefont {Zhong}}]{HeRi2024}%
  \BibitemOpen
  \bibfield  {author} {\bibinfo {author} {\bibfnamefont {R.}~\bibnamefont {He}}, \bibinfo {author} {\bibfnamefont {B.}~\bibnamefont {Zhang}}, \bibinfo {author} {\bibfnamefont {H.}~\bibnamefont {Wang}}, \bibinfo {author} {\bibfnamefont {L.}~\bibnamefont {Li}}, \bibinfo {author} {\bibfnamefont {P.}~\bibnamefont {Tang}}, \bibinfo {author} {\bibfnamefont {G.}~\bibnamefont {Bauer}},\ and\ \bibinfo {author} {\bibfnamefont {Z.}~\bibnamefont {Zhong}},\ }\bibfield  {title} {\bibinfo {title} {Ultrafast switching dynamics of the ferroelectric order in stacking-engineered ferroelectrics},\ }\href {https://doi.org/https://doi.org/10.1016/j.actamat.2023.119416} {\bibfield  {journal} {\bibinfo  {journal} {Acta Materialia}\ }\textbf {\bibinfo {volume} {262}},\ \bibinfo {pages} {119416} (\bibinfo {year} {2024})}\BibitemShut {NoStop}%
\bibitem [{SM()}]{SM}%
  \BibitemOpen
  \href@noop {} {\bibinfo {title} {See supplemental material at ** for {Born} effective charges of {BN} bilayer, computational details, model accuracy, analytical calculation of coercive electric field, and supplemental figures, which include refs.[11,24,35-50]}}\BibitemShut {NoStop}%
\bibitem [{\citenamefont {Kresse}\ and\ \citenamefont {Furthmüller}(1996)}]{Kresse1996}%
  \BibitemOpen
  \bibfield  {author} {\bibinfo {author} {\bibfnamefont {G.}~\bibnamefont {Kresse}}\ and\ \bibinfo {author} {\bibfnamefont {J.}~\bibnamefont {Furthmüller}},\ }\bibfield  {title} {\bibinfo {title} {Efficient iterative schemes for ab initio total-energy calculations using a plane-wave basis set},\ }\href {https://doi.org/10.1103/PhysRevB.54.11169} {\bibfield  {journal} {\bibinfo  {journal} {Physical Review B - Condensed Matter and Materials Physics}\ }\textbf {\bibinfo {volume} {54}},\ \bibinfo {pages} {11169 } (\bibinfo {year} {1996})},\ \bibinfo {note} {cited by: 84293}\BibitemShut {NoStop}%
\bibitem [{\citenamefont {Bl\"ochl}(1994)}]{Blochl1994}%
  \BibitemOpen
  \bibfield  {author} {\bibinfo {author} {\bibfnamefont {P.~E.}\ \bibnamefont {Bl\"ochl}},\ }\bibfield  {title} {\bibinfo {title} {Projector augmented-wave method},\ }\href {https://doi.org/10.1103/PhysRevB.50.17953} {\bibfield  {journal} {\bibinfo  {journal} {Phys. Rev. B}\ }\textbf {\bibinfo {volume} {50}},\ \bibinfo {pages} {17953} (\bibinfo {year} {1994})}\BibitemShut {NoStop}%
\bibitem [{\citenamefont {Perdew}\ \emph {et~al.}(1996)\citenamefont {Perdew}, \citenamefont {Burke},\ and\ \citenamefont {Ernzerhof}}]{Perdew1996}%
  \BibitemOpen
  \bibfield  {author} {\bibinfo {author} {\bibfnamefont {J.~P.}\ \bibnamefont {Perdew}}, \bibinfo {author} {\bibfnamefont {K.}~\bibnamefont {Burke}},\ and\ \bibinfo {author} {\bibfnamefont {M.}~\bibnamefont {Ernzerhof}},\ }\bibfield  {title} {\bibinfo {title} {Generalized gradient approximation made simple},\ }\href {https://doi.org/10.1103/PhysRevLett.77.3865} {\bibfield  {journal} {\bibinfo  {journal} {Phys. Rev. Lett.}\ }\textbf {\bibinfo {volume} {77}},\ \bibinfo {pages} {3865} (\bibinfo {year} {1996})}\BibitemShut {NoStop}%
\bibitem [{\citenamefont {Grimme}(2006)}]{Grimme2006}%
  \BibitemOpen
  \bibfield  {author} {\bibinfo {author} {\bibfnamefont {S.}~\bibnamefont {Grimme}},\ }\bibfield  {title} {\bibinfo {title} {Semiempirical {GGA-type} density functional constructed with a long-range dispersion correction},\ }\href {https://doi.org/https://doi.org/10.1002/jcc.20495} {\bibfield  {journal} {\bibinfo  {journal} {Journal of Computational Chemistry}\ }\textbf {\bibinfo {volume} {27}},\ \bibinfo {pages} {1787} (\bibinfo {year} {2006})},\ \Eprint {https://arxiv.org/abs/https://onlinelibrary.wiley.com/doi/pdf/10.1002/jcc.20495} {https://onlinelibrary.wiley.com/doi/pdf/10.1002/jcc.20495} \BibitemShut {NoStop}%
\bibitem [{\citenamefont {Wang}\ \emph {et~al.}(2019)\citenamefont {Wang}, \citenamefont {Xu}, \citenamefont {Liu}, \citenamefont {Tang},\ and\ \citenamefont {Geng}}]{Vaspkit2019}%
  \BibitemOpen
  \bibfield  {author} {\bibinfo {author} {\bibfnamefont {V.}~\bibnamefont {Wang}}, \bibinfo {author} {\bibfnamefont {N.}~\bibnamefont {Xu}}, \bibinfo {author} {\bibfnamefont {J.~C.}\ \bibnamefont {Liu}}, \bibinfo {author} {\bibfnamefont {G.}~\bibnamefont {Tang}},\ and\ \bibinfo {author} {\bibfnamefont {W.}~\bibnamefont {Geng}},\ }\bibfield  {title} {\bibinfo {title} {{VASPKIT}: A user-friendly interface facilitating high-throughput computing and analysis using vasp code},\ }\href {https://api.semanticscholar.org/CorpusID:220055915} {\bibfield  {journal} {\bibinfo  {journal} {Comput. Phys. Commun.}\ }\textbf {\bibinfo {volume} {267}},\ \bibinfo {pages} {108033} (\bibinfo {year} {2019})}\BibitemShut {NoStop}%
\bibitem [{\citenamefont {Jinnouchi}\ \emph {et~al.}(2019{\natexlab{a}})\citenamefont {Jinnouchi}, \citenamefont {Lahnsteiner}, \citenamefont {Karsai}, \citenamefont {Kresse},\ and\ \citenamefont {Bokdam}}]{Jinnouchi2019}%
  \BibitemOpen
  \bibfield  {author} {\bibinfo {author} {\bibfnamefont {R.}~\bibnamefont {Jinnouchi}}, \bibinfo {author} {\bibfnamefont {J.}~\bibnamefont {Lahnsteiner}}, \bibinfo {author} {\bibfnamefont {F.}~\bibnamefont {Karsai}}, \bibinfo {author} {\bibfnamefont {G.}~\bibnamefont {Kresse}},\ and\ \bibinfo {author} {\bibfnamefont {M.}~\bibnamefont {Bokdam}},\ }\bibfield  {title} {\bibinfo {title} {Phase transitions of hybrid perovskites simulated by machine-learning force fields trained on the fly with {Bayesian} inference},\ }\href {https://doi.org/10.1103/PhysRevLett.122.225701} {\bibfield  {journal} {\bibinfo  {journal} {Phys. Rev. Lett.}\ }\textbf {\bibinfo {volume} {122}},\ \bibinfo {pages} {225701} (\bibinfo {year} {2019}{\natexlab{a}})}\BibitemShut {NoStop}%
\bibitem [{\citenamefont {Jinnouchi}\ \emph {et~al.}(2019{\natexlab{b}})\citenamefont {Jinnouchi}, \citenamefont {Karsai},\ and\ \citenamefont {Kresse}}]{Jinnouchi2019_2}%
  \BibitemOpen
  \bibfield  {author} {\bibinfo {author} {\bibfnamefont {R.}~\bibnamefont {Jinnouchi}}, \bibinfo {author} {\bibfnamefont {F.}~\bibnamefont {Karsai}},\ and\ \bibinfo {author} {\bibfnamefont {G.}~\bibnamefont {Kresse}},\ }\bibfield  {title} {\bibinfo {title} {On-the-fly machine learning force field generation: Application to melting points},\ }\href {https://doi.org/10.1103/PhysRevB.100.014105} {\bibfield  {journal} {\bibinfo  {journal} {Phys. Rev. B}\ }\textbf {\bibinfo {volume} {100}},\ \bibinfo {pages} {014105} (\bibinfo {year} {2019}{\natexlab{b}})}\BibitemShut {NoStop}%
\bibitem [{\citenamefont {Jinnouchi}\ \emph {et~al.}(2020)\citenamefont {Jinnouchi}, \citenamefont {Karsai}, \citenamefont {Verdi}, \citenamefont {Asahi},\ and\ \citenamefont {Kresse}}]{Jinnouchi2020}%
  \BibitemOpen
  \bibfield  {author} {\bibinfo {author} {\bibfnamefont {R.}~\bibnamefont {Jinnouchi}}, \bibinfo {author} {\bibfnamefont {F.}~\bibnamefont {Karsai}}, \bibinfo {author} {\bibfnamefont {C.}~\bibnamefont {Verdi}}, \bibinfo {author} {\bibfnamefont {R.}~\bibnamefont {Asahi}},\ and\ \bibinfo {author} {\bibfnamefont {G.}~\bibnamefont {Kresse}},\ }\bibfield  {title} {\bibinfo {title} {{Descriptors representing two- and three-body atomic distributions and their effects on the accuracy of machine-learned inter-atomic potentials}},\ }\href {https://doi.org/10.1063/5.0009491} {\bibfield  {journal} {\bibinfo  {journal} {The Journal of Chemical Physics}\ }\textbf {\bibinfo {volume} {152}},\ \bibinfo {pages} {234102} (\bibinfo {year} {2020})},\ \Eprint {https://arxiv.org/abs/https://pubs.aip.org/aip/jcp/article-pdf/doi/10.1063/5.0009491/15575269/234102\_1\_online.pdf} {https://pubs.aip.org/aip/jcp/article-pdf/doi/10.1063/5.0009491/15575269/234102\_1\_online.pdf} \BibitemShut {NoStop}%
\bibitem [{\citenamefont {Allen}\ and\ \citenamefont {Tildesley}(2017)}]{Allen2017}%
  \BibitemOpen
  \bibfield  {author} {\bibinfo {author} {\bibfnamefont {M.~P.}\ \bibnamefont {Allen}}\ and\ \bibinfo {author} {\bibfnamefont {D.~J.}\ \bibnamefont {Tildesley}},\ }\href {https://doi.org/10.1093/oso/9780198803195.001.0001} {\emph {\bibinfo {title} {{Computer Simulation of Liquids}}}}\ (\bibinfo  {publisher} {Oxford University Press},\ \bibinfo {year} {2017})\BibitemShut {NoStop}%
\bibitem [{\citenamefont {Gonze}\ and\ \citenamefont {Lee}(1997)}]{Gonze1997}%
  \BibitemOpen
  \bibfield  {author} {\bibinfo {author} {\bibfnamefont {X.}~\bibnamefont {Gonze}}\ and\ \bibinfo {author} {\bibfnamefont {C.}~\bibnamefont {Lee}},\ }\bibfield  {title} {\bibinfo {title} {Dynamical matrices, {Born} effective charges, dielectric permittivity tensors, and interatomic force constants from density-functional perturbation theory},\ }\href {https://doi.org/10.1103/PhysRevB.55.10355} {\bibfield  {journal} {\bibinfo  {journal} {Phys. Rev. B}\ }\textbf {\bibinfo {volume} {55}},\ \bibinfo {pages} {10355} (\bibinfo {year} {1997})}\BibitemShut {NoStop}%
\bibitem [{\citenamefont {Musaelian}\ \emph {et~al.}(2023)\citenamefont {Musaelian}, \citenamefont {Batzner}, \citenamefont {Johansson}, \citenamefont {Sun}, \citenamefont {Owen}, \citenamefont {Kornbluth},\ and\ \citenamefont {Kozinsky}}]{Musaelian2023}%
  \BibitemOpen
  \bibfield  {author} {\bibinfo {author} {\bibfnamefont {A.}~\bibnamefont {Musaelian}}, \bibinfo {author} {\bibfnamefont {S.}~\bibnamefont {Batzner}}, \bibinfo {author} {\bibfnamefont {A.}~\bibnamefont {Johansson}}, \bibinfo {author} {\bibfnamefont {L.}~\bibnamefont {Sun}}, \bibinfo {author} {\bibfnamefont {C.~J.}\ \bibnamefont {Owen}}, \bibinfo {author} {\bibfnamefont {M.}~\bibnamefont {Kornbluth}},\ and\ \bibinfo {author} {\bibfnamefont {B.}~\bibnamefont {Kozinsky}},\ }\bibfield  {title} {\bibinfo {title} {Learning local equivariant representations for large-scale atomistic dynamics},\ }\href {https://doi.org/10.1038/s41467-023-36329-y} {\bibfield  {journal} {\bibinfo  {journal} {Nature Communications}\ }\textbf {\bibinfo {volume} {14}},\ \bibinfo {pages} {579} (\bibinfo {year} {2023})}\BibitemShut {NoStop}%
\bibitem [{\citenamefont {Yu}\ \emph {et~al.}(2024)\citenamefont {Yu}, \citenamefont {Deng}, \citenamefont {Xie}, \citenamefont {Zhang}, \citenamefont {Shi}, \citenamefont {Zhong}, \citenamefont {He},\ and\ \citenamefont {Xiang}}]{YuHongyu2024}%
  \BibitemOpen
  \bibfield  {author} {\bibinfo {author} {\bibfnamefont {H.}~\bibnamefont {Yu}}, \bibinfo {author} {\bibfnamefont {S.}~\bibnamefont {Deng}}, \bibinfo {author} {\bibfnamefont {M.}~\bibnamefont {Xie}}, \bibinfo {author} {\bibfnamefont {Y.}~\bibnamefont {Zhang}}, \bibinfo {author} {\bibfnamefont {X.}~\bibnamefont {Shi}}, \bibinfo {author} {\bibfnamefont {J.}~\bibnamefont {Zhong}}, \bibinfo {author} {\bibfnamefont {C.}~\bibnamefont {He}},\ and\ \bibinfo {author} {\bibfnamefont {H.}~\bibnamefont {Xiang}},\ }\href {https://arxiv.org/abs/2407.01914} {\bibinfo {title} {Switchable ferroelectricity in subnano silicon thin films}} (\bibinfo {year} {2024}),\ \Eprint {https://arxiv.org/abs/2407.01914} {arXiv:2407.01914 [cond-mat.mtrl-sci]} \BibitemShut {NoStop}%
\bibitem [{\citenamefont {Thompson}\ \emph {et~al.}(2022)\citenamefont {Thompson}, \citenamefont {Aktulga}, \citenamefont {Berger}, \citenamefont {Bolintineanu}, \citenamefont {Brown}, \citenamefont {Crozier}, \citenamefont {{in 't Veld}}, \citenamefont {Kohlmeyer}, \citenamefont {Moore}, \citenamefont {Nguyen}, \citenamefont {Shan}, \citenamefont {Stevens}, \citenamefont {Tranchida}, \citenamefont {Trott},\ and\ \citenamefont {Plimpton}}]{LAMMPS2022}%
  \BibitemOpen
  \bibfield  {author} {\bibinfo {author} {\bibfnamefont {A.~P.}\ \bibnamefont {Thompson}}, \bibinfo {author} {\bibfnamefont {H.~M.}\ \bibnamefont {Aktulga}}, \bibinfo {author} {\bibfnamefont {R.}~\bibnamefont {Berger}}, \bibinfo {author} {\bibfnamefont {D.~S.}\ \bibnamefont {Bolintineanu}}, \bibinfo {author} {\bibfnamefont {W.~M.}\ \bibnamefont {Brown}}, \bibinfo {author} {\bibfnamefont {P.~S.}\ \bibnamefont {Crozier}}, \bibinfo {author} {\bibfnamefont {P.~J.}\ \bibnamefont {{in 't Veld}}}, \bibinfo {author} {\bibfnamefont {A.}~\bibnamefont {Kohlmeyer}}, \bibinfo {author} {\bibfnamefont {S.~G.}\ \bibnamefont {Moore}}, \bibinfo {author} {\bibfnamefont {T.~D.}\ \bibnamefont {Nguyen}}, \bibinfo {author} {\bibfnamefont {R.}~\bibnamefont {Shan}}, \bibinfo {author} {\bibfnamefont {M.~J.}\ \bibnamefont {Stevens}}, \bibinfo {author} {\bibfnamefont {J.}~\bibnamefont {Tranchida}}, \bibinfo {author} {\bibfnamefont {C.}~\bibnamefont {Trott}},\ and\ \bibinfo {author} {\bibfnamefont {S.~J.}\ \bibnamefont {Plimpton}},\ }\bibfield  {title} {\bibinfo {title} {Lammps - a flexible simulation tool for particle-based materials modeling at the atomic, meso, and continuum scales},\ }\href {https://doi.org/https://doi.org/10.1016/j.cpc.2021.108171} {\bibfield  {journal} {\bibinfo  {journal} {Computer Physics Communications}\ }\textbf {\bibinfo {volume} {271}},\ \bibinfo {pages} {108171} (\bibinfo {year} {2022})}\BibitemShut {NoStop}%
\bibitem [{\citenamefont {Nosé}(1984)}]{Nosé1984}%
  \BibitemOpen
  \bibfield  {author} {\bibinfo {author} {\bibfnamefont {S.}~\bibnamefont {Nosé}},\ }\bibfield  {title} {\bibinfo {title} {A unified formulation of the constant temperature molecular dynamics methods},\ }\href {https://doi.org/10.1063/1.447334} {\bibfield  {journal} {\bibinfo  {journal} {The Journal of Chemical Physics}\ }\textbf {\bibinfo {volume} {81}},\ \bibinfo {pages} {511 } (\bibinfo {year} {1984})},\ \bibinfo {note} {cited by: 13724; All Open Access, Green Open Access}\BibitemShut {NoStop}%
\bibitem [{\citenamefont {{OriginLab Corporation}}(2024)}]{origin2024}%
  \BibitemOpen
  \bibfield  {author} {\bibinfo {author} {\bibnamefont {{OriginLab Corporation}}},\ }\href@noop {} {\bibinfo {title} {{Origin}}} (\bibinfo {year} {2024})\BibitemShut {NoStop}%
\bibitem [{\citenamefont {Weston}\ \emph {et~al.}(2020)\citenamefont {Weston}, \citenamefont {Zou}, \citenamefont {Enaldiev}, \citenamefont {Summerfield}, \citenamefont {Clark}, \citenamefont {Z{\'o}lyomi}, \citenamefont {Graham}, \citenamefont {Yelgel}, \citenamefont {Magorrian}, \citenamefont {Zhou}, \citenamefont {Zultak}, \citenamefont {Hopkinson}, \citenamefont {Barinov}, \citenamefont {Bointon}, \citenamefont {Kretinin}, \citenamefont {Wilson}, \citenamefont {Beton}, \citenamefont {Fal'ko}, \citenamefont {Haigh},\ and\ \citenamefont {Gorbachev}}]{Weston2020}%
  \BibitemOpen
  \bibfield  {author} {\bibinfo {author} {\bibfnamefont {A.}~\bibnamefont {Weston}}, \bibinfo {author} {\bibfnamefont {Y.}~\bibnamefont {Zou}}, \bibinfo {author} {\bibfnamefont {V.}~\bibnamefont {Enaldiev}}, \bibinfo {author} {\bibfnamefont {A.}~\bibnamefont {Summerfield}}, \bibinfo {author} {\bibfnamefont {N.}~\bibnamefont {Clark}}, \bibinfo {author} {\bibfnamefont {V.}~\bibnamefont {Z{\'o}lyomi}}, \bibinfo {author} {\bibfnamefont {A.}~\bibnamefont {Graham}}, \bibinfo {author} {\bibfnamefont {C.}~\bibnamefont {Yelgel}}, \bibinfo {author} {\bibfnamefont {S.}~\bibnamefont {Magorrian}}, \bibinfo {author} {\bibfnamefont {M.}~\bibnamefont {Zhou}}, \bibinfo {author} {\bibfnamefont {J.}~\bibnamefont {Zultak}}, \bibinfo {author} {\bibfnamefont {D.}~\bibnamefont {Hopkinson}}, \bibinfo {author} {\bibfnamefont {A.}~\bibnamefont {Barinov}}, \bibinfo {author} {\bibfnamefont {T.~H.}\ \bibnamefont {Bointon}}, \bibinfo {author} {\bibfnamefont {A.}~\bibnamefont {Kretinin}}, \bibinfo {author} {\bibfnamefont {N.~R.}\ \bibnamefont {Wilson}}, \bibinfo {author} {\bibfnamefont {P.~H.}\ \bibnamefont {Beton}}, \bibinfo {author} {\bibfnamefont {V.~I.}\ \bibnamefont {Fal'ko}}, \bibinfo {author} {\bibfnamefont {S.~J.}\ \bibnamefont {Haigh}},\ and\ \bibinfo {author} {\bibfnamefont {R.}~\bibnamefont {Gorbachev}},\ }\bibfield  {title} {\bibinfo {title} {Atomic reconstruction in twisted bilayers of transition metal dichalcogenides},\ }\href {https://doi.org/10.1038/s41565-020-0682-9} {\bibfield  {journal} {\bibinfo  {journal} {Nature Nanotechnology}\ }\textbf {\bibinfo {volume} {15}},\ \bibinfo {pages} {592} (\bibinfo {year} {2020})}\BibitemShut {NoStop}%
\bibitem [{\citenamefont {Shi}\ \emph {et~al.}(2023)\citenamefont {Shi}, \citenamefont {Wang}, \citenamefont {Jiang}, \citenamefont {Feng}, \citenamefont {Guo}, \citenamefont {Gao}, \citenamefont {Gao},\ and\ \citenamefont {Ren}}]{ShiBowen2023}%
  \BibitemOpen
  \bibfield  {author} {\bibinfo {author} {\bibfnamefont {B.}~\bibnamefont {Shi}}, \bibinfo {author} {\bibfnamefont {H.}~\bibnamefont {Wang}}, \bibinfo {author} {\bibfnamefont {W.}~\bibnamefont {Jiang}}, \bibinfo {author} {\bibfnamefont {Y.}~\bibnamefont {Feng}}, \bibinfo {author} {\bibfnamefont {P.}~\bibnamefont {Guo}}, \bibinfo {author} {\bibfnamefont {H.}~\bibnamefont {Gao}}, \bibinfo {author} {\bibfnamefont {Z.}~\bibnamefont {Gao}},\ and\ \bibinfo {author} {\bibfnamefont {W.}~\bibnamefont {Ren}},\ }\bibfield  {title} {\bibinfo {title} {Electric field effect of sliding graphene/hexagonal boron nitride heterobilayer},\ }\href {https://doi.org/https://doi.org/10.1016/j.apsusc.2023.157816} {\bibfield  {journal} {\bibinfo  {journal} {Applied Surface Science}\ }\textbf {\bibinfo {volume} {636}},\ \bibinfo {pages} {157816} (\bibinfo {year} {2023})}\BibitemShut {NoStop}%
\bibitem [{\citenamefont {King-Smith}\ and\ \citenamefont {Vanderbilt}(1993)}]{KingSmith1993}%
  \BibitemOpen
  \bibfield  {author} {\bibinfo {author} {\bibfnamefont {R.~D.}\ \bibnamefont {King-Smith}}\ and\ \bibinfo {author} {\bibfnamefont {D.}~\bibnamefont {Vanderbilt}},\ }\bibfield  {title} {\bibinfo {title} {Theory of polarization of crystalline solids},\ }\href {https://doi.org/10.1103/PhysRevB.47.1651} {\bibfield  {journal} {\bibinfo  {journal} {Phys. Rev. B}\ }\textbf {\bibinfo {volume} {47}},\ \bibinfo {pages} {1651} (\bibinfo {year} {1993})}\BibitemShut {NoStop}%
\bibitem [{\citenamefont {Resta}(1992)}]{Resta1992}%
  \BibitemOpen
  \bibfield  {author} {\bibinfo {author} {\bibfnamefont {R.}~\bibnamefont {Resta}},\ }\bibfield  {title} {\bibinfo {title} {Theory of the electric polarization in crystals},\ }\href {https://doi.org/10.1080/00150199208016065} {\bibfield  {journal} {\bibinfo  {journal} {Ferroelectrics}\ }\textbf {\bibinfo {volume} {136}},\ \bibinfo {pages} {51} (\bibinfo {year} {1992})},\ \Eprint {https://arxiv.org/abs/https://doi.org/10.1080/00150199208016065} {https://doi.org/10.1080/00150199208016065} \BibitemShut {NoStop}%
\bibitem [{\citenamefont {Batzner}\ \emph {et~al.}(2022)\citenamefont {Batzner}, \citenamefont {Musaelian}, \citenamefont {Sun}, \citenamefont {Geiger}, \citenamefont {Mailoa}, \citenamefont {Kornbluth}, \citenamefont {Molinari}, \citenamefont {Smidt},\ and\ \citenamefont {Kozinsky}}]{Batzner2022}%
  \BibitemOpen
  \bibfield  {author} {\bibinfo {author} {\bibfnamefont {S.}~\bibnamefont {Batzner}}, \bibinfo {author} {\bibfnamefont {A.}~\bibnamefont {Musaelian}}, \bibinfo {author} {\bibfnamefont {L.}~\bibnamefont {Sun}}, \bibinfo {author} {\bibfnamefont {M.}~\bibnamefont {Geiger}}, \bibinfo {author} {\bibfnamefont {J.~P.}\ \bibnamefont {Mailoa}}, \bibinfo {author} {\bibfnamefont {M.}~\bibnamefont {Kornbluth}}, \bibinfo {author} {\bibfnamefont {N.}~\bibnamefont {Molinari}}, \bibinfo {author} {\bibfnamefont {T.~E.}\ \bibnamefont {Smidt}},\ and\ \bibinfo {author} {\bibfnamefont {B.}~\bibnamefont {Kozinsky}},\ }\bibfield  {title} {\bibinfo {title} {E(3)-equivariant graph neural networks for data-efficient and accurate interatomic potentials},\ }\href {https://doi.org/10.1038/s41467-022-29939-5} {\bibfield  {journal} {\bibinfo  {journal} {Nature Communications}\ }\textbf {\bibinfo {volume} {13}},\ \bibinfo {pages} {2453} (\bibinfo {year} {2022})}\BibitemShut {NoStop}%
\bibitem [{\citenamefont {Gasteiger}\ \emph {et~al.}(2022)\citenamefont {Gasteiger}, \citenamefont {Groß},\ and\ \citenamefont {Günnemann}}]{Gasteiger2022}%
  \BibitemOpen
  \bibfield  {author} {\bibinfo {author} {\bibfnamefont {J.}~\bibnamefont {Gasteiger}}, \bibinfo {author} {\bibfnamefont {J.}~\bibnamefont {Groß}},\ and\ \bibinfo {author} {\bibfnamefont {S.}~\bibnamefont {Günnemann}},\ }\href@noop {} {\bibinfo {title} {Directional message passing for molecular graphs}} (\bibinfo {year} {2022}),\ \Eprint {https://arxiv.org/abs/2003.03123} {arXiv:2003.03123 [cs.LG]} \BibitemShut {NoStop}%
\bibitem [{\citenamefont {Landauer}(1957)}]{Landauer1957}%
  \BibitemOpen
  \bibfield  {author} {\bibinfo {author} {\bibfnamefont {R.}~\bibnamefont {Landauer}},\ }\bibfield  {title} {\bibinfo {title} {Electrostatic considerations in {BaTiO3} domain formation during polarization reversal},\ }\href {https://doi.org/10.1063/1.1722712} {\bibfield  {journal} {\bibinfo  {journal} {Journal of Applied Physics}\ }\textbf {\bibinfo {volume} {28}},\ \bibinfo {pages} {227} (\bibinfo {year} {1957})},\ \Eprint {https://arxiv.org/abs/https://pubs.aip.org/aip/jap/article-pdf/28/2/227/18316454/227\_1\_online.pdf} {https://pubs.aip.org/aip/jap/article-pdf/28/2/227/18316454/227\_1\_online.pdf} \BibitemShut {NoStop}%
\bibitem [{\citenamefont {Xu}\ \emph {et~al.}(2017{\natexlab{a}})\citenamefont {Xu}, \citenamefont {Garcia}, \citenamefont {Fusil}, \citenamefont {Bibes},\ and\ \citenamefont {Bellaiche}}]{XuBin2017}%
  \BibitemOpen
  \bibfield  {author} {\bibinfo {author} {\bibfnamefont {B.}~\bibnamefont {Xu}}, \bibinfo {author} {\bibfnamefont {V.}~\bibnamefont {Garcia}}, \bibinfo {author} {\bibfnamefont {S.}~\bibnamefont {Fusil}}, \bibinfo {author} {\bibfnamefont {M.}~\bibnamefont {Bibes}},\ and\ \bibinfo {author} {\bibfnamefont {L.}~\bibnamefont {Bellaiche}},\ }\bibfield  {title} {\bibinfo {title} {Intrinsic polarization switching mechanisms in {${\mathrm{BiFeO}}_{3}$}},\ }\href {https://doi.org/10.1103/PhysRevB.95.104104} {\bibfield  {journal} {\bibinfo  {journal} {Phys. Rev. B}\ }\textbf {\bibinfo {volume} {95}},\ \bibinfo {pages} {104104} (\bibinfo {year} {2017}{\natexlab{a}})}\BibitemShut {NoStop}%
\bibitem [{\citenamefont {Xu}\ \emph {et~al.}(2017{\natexlab{b}})\citenamefont {Xu}, \citenamefont {{\'I}{\~{n}}iguez},\ and\ \citenamefont {Bellaiche}}]{XuBin2017_2}%
  \BibitemOpen
  \bibfield  {author} {\bibinfo {author} {\bibfnamefont {B.}~\bibnamefont {Xu}}, \bibinfo {author} {\bibfnamefont {J.}~\bibnamefont {{\'I}{\~{n}}iguez}},\ and\ \bibinfo {author} {\bibfnamefont {L.}~\bibnamefont {Bellaiche}},\ }\bibfield  {title} {\bibinfo {title} {Designing lead-free antiferroelectrics for energy storage},\ }\href {https://doi.org/10.1038/ncomms15682} {\bibfield  {journal} {\bibinfo  {journal} {Nature Communications}\ }\textbf {\bibinfo {volume} {8}},\ \bibinfo {pages} {15682} (\bibinfo {year} {2017}{\natexlab{b}})}\BibitemShut {NoStop}%
\bibitem [{\citenamefont {Daumont}\ \emph {et~al.}(2012)\citenamefont {Daumont}, \citenamefont {Ren}, \citenamefont {Infante}, \citenamefont {Lisenkov}, \citenamefont {Allibe}, \citenamefont {Carrétéro}, \citenamefont {Fusil}, \citenamefont {Jacquet}, \citenamefont {Bouvet}, \citenamefont {Bouamrane}, \citenamefont {Prosandeev}, \citenamefont {Geneste}, \citenamefont {Dkhil}, \citenamefont {Bellaiche}, \citenamefont {Barthélémy},\ and\ \citenamefont {Bibes}}]{Daumont2012}%
  \BibitemOpen
  \bibfield  {author} {\bibinfo {author} {\bibfnamefont {C.}~\bibnamefont {Daumont}}, \bibinfo {author} {\bibfnamefont {W.}~\bibnamefont {Ren}}, \bibinfo {author} {\bibfnamefont {I.~C.}\ \bibnamefont {Infante}}, \bibinfo {author} {\bibfnamefont {S.}~\bibnamefont {Lisenkov}}, \bibinfo {author} {\bibfnamefont {J.}~\bibnamefont {Allibe}}, \bibinfo {author} {\bibfnamefont {C.}~\bibnamefont {Carrétéro}}, \bibinfo {author} {\bibfnamefont {S.}~\bibnamefont {Fusil}}, \bibinfo {author} {\bibfnamefont {E.}~\bibnamefont {Jacquet}}, \bibinfo {author} {\bibfnamefont {T.}~\bibnamefont {Bouvet}}, \bibinfo {author} {\bibfnamefont {F.}~\bibnamefont {Bouamrane}}, \bibinfo {author} {\bibfnamefont {S.}~\bibnamefont {Prosandeev}}, \bibinfo {author} {\bibfnamefont {G.}~\bibnamefont {Geneste}}, \bibinfo {author} {\bibfnamefont {B.}~\bibnamefont {Dkhil}}, \bibinfo {author} {\bibfnamefont {L.}~\bibnamefont {Bellaiche}}, \bibinfo {author} {\bibfnamefont {A.}~\bibnamefont {Barthélémy}},\ and\ \bibinfo {author} {\bibfnamefont {M.}~\bibnamefont {Bibes}},\ }\bibfield  {title} {\bibinfo {title} {Strain dependence of polarization and piezoelectric response in epitaxial {BiFeO3} thin films},\ }\href {https://doi.org/10.1088/0953-8984/24/16/162202} {\bibfield  {journal} {\bibinfo  {journal} {Journal of Physics: Condensed Matter}\ }\textbf {\bibinfo {volume} {24}},\ \bibinfo {pages} {162202} (\bibinfo {year} {2012})}\BibitemShut {NoStop}%
\bibitem [{\citenamefont {Lisenkov}\ \emph {et~al.}(2009)\citenamefont {Lisenkov}, \citenamefont {Rahmedov},\ and\ \citenamefont {Bellaiche}}]{Lisenkov2009}%
  \BibitemOpen
  \bibfield  {author} {\bibinfo {author} {\bibfnamefont {S.}~\bibnamefont {Lisenkov}}, \bibinfo {author} {\bibfnamefont {D.}~\bibnamefont {Rahmedov}},\ and\ \bibinfo {author} {\bibfnamefont {L.}~\bibnamefont {Bellaiche}},\ }\bibfield  {title} {\bibinfo {title} {Electric-field-induced paths in multiferroic {${\mathrm{BiFeO}}_{3}$} from atomistic simulations},\ }\href {https://doi.org/10.1103/PhysRevLett.103.047204} {\bibfield  {journal} {\bibinfo  {journal} {Phys. Rev. Lett.}\ }\textbf {\bibinfo {volume} {103}},\ \bibinfo {pages} {047204} (\bibinfo {year} {2009})}\BibitemShut {NoStop}%
\end{thebibliography}%

\newpage

\titleformat{\section}[block]
  {\normalfont\bfseries}{\thesection}{1em}{}
\setcounter{secnumdepth}{0}

\renewcommand{\thefigure}{S\arabic{figure}}
\setcounter{figure}{0}
\renewcommand{\theequation}{S\arabic{equation}}

\section{S\uppercase\expandafter{\romannumeral1}. Born effective charges (BECs) of BN bilayer}
\begin{figure}[h]
    \includegraphics[width=0.9\textwidth]{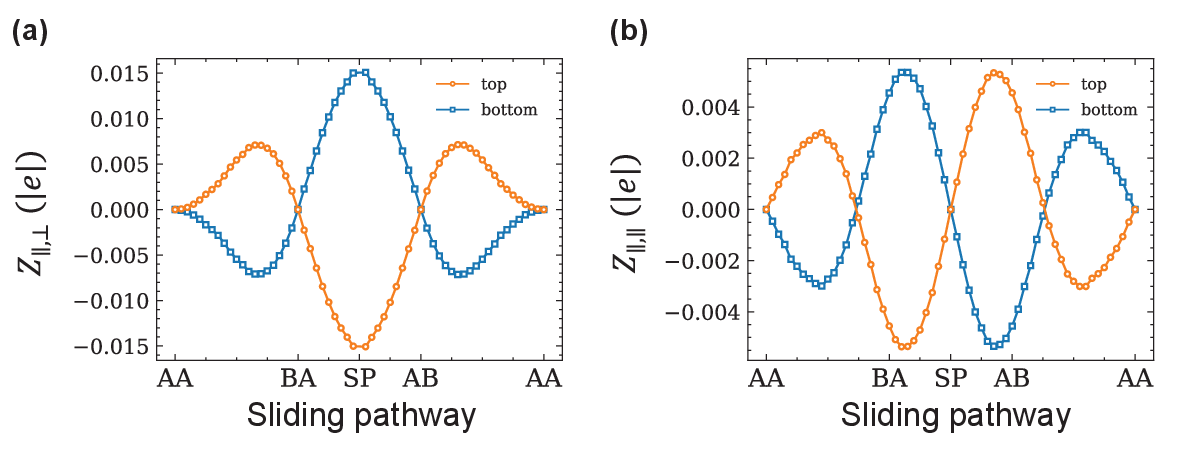}
    \caption{\label{fig:S1-BEC} The sum of sliding-related BEC components of adjacent $\mathrm{B}$ and $\mathrm{N}$ ions in the top layer (orange line) and bottom layer (blue line) along the sliding pathway, calculated using DFPT. (a) The BEC component related to $E_{\perp}$ and the in-plane force in the sliding direction $F_{\parallel}$ ($F_{\parallel} = Z_{\rm \parallel, \perp}E_{\perp}$). (b) The BEC component related to $E_{\parallel}$ and $F_{\parallel}$ ($F_{\parallel} = Z_{\parallel, \parallel}E_{\parallel}$).}
\end{figure}

    The force acting on ion $i$ in the presence of a finite electric field $E$ can be determined using the formula $\bm{F}_i=Z_i(\bm{R})\bm{E}$, where $Z_i(\bm{R})$ denotes the zero-field Born effective change (BEC) tensor of ion $i$ at the coordinates $\bm{R}$. During polarization switching in sliding ferroelectrics, that the positional rearrangement among ions within the same layer remains negligible, with the primary change attributed to interlayer sliding. Consequently, we group two adjacent $\mathrm{B}$ and $\mathrm{N}$ ions within the same layer as a collective entity to examine the BECs distribution along the sliding pathway. Illustratively, as depicted in Fig.~\ref{fig:S1-BEC}, ions experience forces of equal magnitude but opposite direction under the influence of an external field. Specifically, when $E_{\perp}$ is applied, the distribution of forces is symmetric with respect to the transition state SP [see Fig.~\ref{fig:S1-BEC}(a)]. Conversely, with $E_{parallel}$ applied in the sliding direction, the distribution becomes antisymmetric [see Fig.~\ref{fig:S1-BEC}(b)]. In both situations, the BECs exhibit significant variations along the sliding path from BA to AB. Previous MD simulations that use fixed BEC values cannot precisely capture the behavior of the $\mathrm{BN}$ bilayer under these electric field conditions. This limitation is particularly evident for $E_{parallel}$, where it cannot calculate the force by employing the average BEC value along the sliding pathway, which would be zero. Hence, our proposed model, capable of predicting BECs for varying coordinates $\bm{R}$, proves to be indispensable.

\section{S\uppercase\expandafter{\romannumeral2}. Computational details and model accuracy}
    We employ the Vienna \emph{ab initio} Simulation Package (VASP) \cite{Kresse1996} to perform DFT calculations using the projector augmented wave (PAW) method \cite{Blochl1994} and the generalized gradient approximation (GGA) with the Perdew-Burke-Ernzerh (PBE) exchange-correlation functional \cite{Perdew1996}. The plane-wave cutoff energy is set to 520 eV. The cell constant along the $c$-axis is fixed at 30 {\AA}, ensuring a sufficient vacuum space. A $12\times12\times1$ $k$-point mesh and vdW corrections with the DFT-D2 method of Grimme \cite{Grimme2006} are applied for structure optimization. The optimized lattice constant of the AB-stacked $\mathrm{BN}$ bilayer is $a=b=2.511$ {\AA} with an interlayer distance of $3.074$ {\AA}, which is consistent with previous theoretical calculations \cite{JiangWen2022}. For other parallel-stacked configurations corresponding to different sliding vectors, the lattice constant remains almost unchanged, and the interlayer distance increases slightly (not exceeding 0.35 {\AA}). The primary cells are expanded to $5\times5\times1$ supercells using Vaspkit \cite{Vaspkit2019}, and MD calculations are performed using on-the-fly machine learning force fields (MLFF) \cite{Jinnouchi2019, Jinnouchi2019_2, Jinnouchi2020} for more efficient sampling. We use a $3\times3\times1$ $k$-point mesh and the NVT ensemble with a Langevin thermostat \cite{Allen2017}. The BECs of each atom in different configurations are further calculated using density functional perturbation theory (DFPT) \cite{Gonze1997}.

\begin{figure}[h]
    \includegraphics[width=0.9\textwidth]{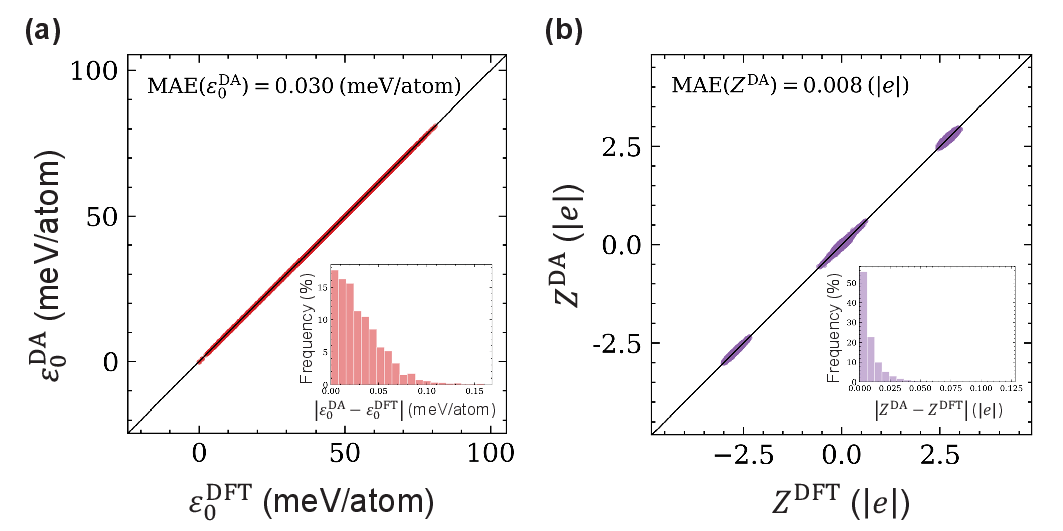}
    \caption{\label{fig:S2-MAE} The parity plot of the DREAM-Allegro (DA) model compared to DFT results. Comparison of energies $\varepsilon_0$ (a) and BECs $Z$ (b) from the DA model against DFT calculations for all configurations in the training dataset. The insets display the distribution of the MAEs.}
\end{figure}

    In our training, the cutoff radius is set to $R_c=7\text{\r{A}}$, encompassing nearly 100 pairs of each atom formed within the 19th nearest neighbor. The learning rate is designed to decay exponentially from $5\times10^{-3}$ to $9.7\times10^{-6}$. For the Allegro \cite{Musaelian2023} model, trained on a dataset containing 1659 structures from MD sampling blow 3000K, we select a (64, 128, 256, 256) dimensional embedding network, a (256, 256, 256) dimensional fitting network, and a (128) dimensional per-edge final network. This configuration yields mean absolute errors (MAEs) of 0.053 meV/atom for energy, 5.311 meV/{\AA} for forces and 0.063 meV/{\AA}$^3$ for stress. For our DREAM-Allegro \cite{YuHongyu2024} model (DA), trained on a dataset containing 3484 structures from MD sampling blow 500K, the loss weight coefficients for energy, forces, stress, and BEC$_0$ (which measures the system's electrical neutrality) are all set to 1, and the coefficient for BEC is increased to 10 due to the relatively small absolute values of BECs. The per-edge final network dimensions are adjusted to (128, 128, 64) dimensions, resulting in a model with MAEs of 0.030 meV/atom for energy, 3.851 meV/{\AA} for forces, 0.076 meV/{\AA}$^3$ for stress and 0.008 $|$e$|$ for BEC (after correcting with BEC$_0$). This indicates that our models are highly accurate. Fig.~\ref{fig:S2-MAE} presents the parity plot of energy and BEC.
    
    The MD simulations are performed using the LAMMPS software package \cite{LAMMPS2022} with a newly integrated atom type named ``BEC''. Periodic boundary conditions are applied along the $a$ and $b$ axes, while a non-periodic, fixed boundary condition is used along the $c$ axis to accurately represent 2D systems. The canonical NVT ensemble, with temperature regulation provided by a Nose-Hoover thermostat \cite{Nosé1984}, is employed. The simulation box is a $15\times15\times1$ supercell, consisting of 900 atoms, with an integration time step of 1 femtosecond (fs).

\section{S\uppercase\expandafter{\romannumeral3}. Analytical calculation of coercive electric field}
\begin{figure}[h]
    \includegraphics[width=\textwidth]{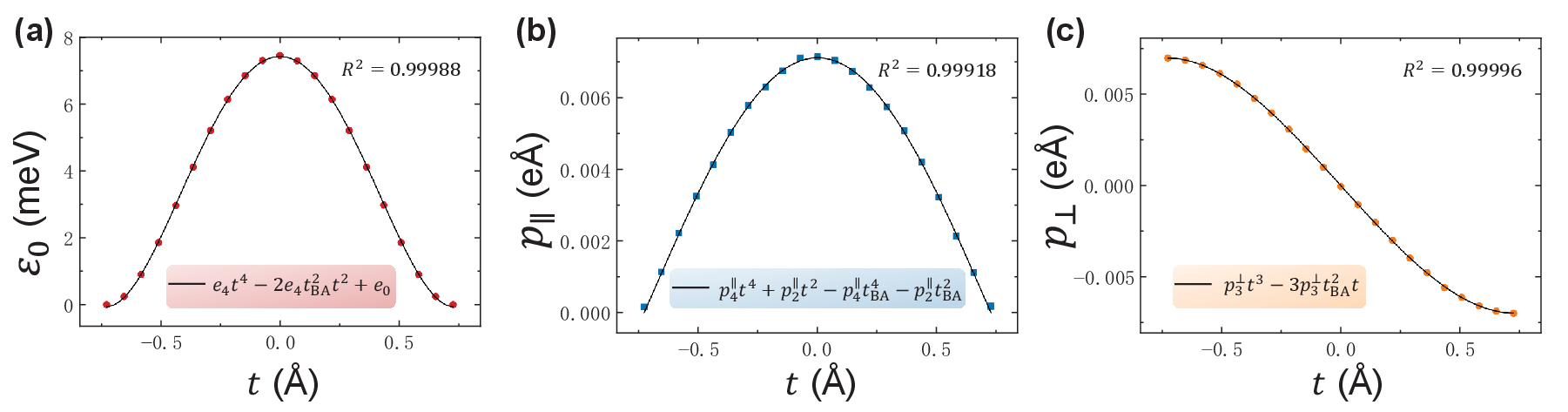}
    \caption{\label{fig:S3-fitting} Fit plots and goodness of fit ($R^2$) for energies $\varepsilon_0(t)$ (a), in-plane dipoles $p_{\parallel}(t)$ (b) and out-of-plane dipoles $p_{\perp}(t)$ (c) per unit cell.}
\end{figure}

    The fitting of energies and dipoles per unit cell along the BA to AB sliding pathway is performed using Origin \cite{origin2024}. Considering their even and odd function characteristics, $\varepsilon_0(t)$ and $p_{\parallel}(t)$ are fitted with quartic polynomials, and $p_{\perp}(t)$ is fitted with cubic polynomials.Constraints are applied to ensure that $\varepsilon_0$ is minimized at the BA (AB) state, $p_{\parallel}$ is zero at the BA (AB) state, and $p_{\perp}$ reaches a maximum (minimum) at the BA (AB) state. The fitting results are presented in Fig.\ref{fig:S3-fitting}. 

    When the critical electric field is defined as the minimum field at which the local energy minimum of $\varepsilon(t)$ disappears, i.e., $\mathrm{d}\varepsilon(t)/\mathrm{d}t \leq 0$ along the BA to SP sliding pathway, the exact analytical expression for $E_{\perp, c}$ is given by:
    \begin{gather}
        \label{eqn:exact_Eperp}
        E_{\perp, c} = \frac{-2 e_4 t_{\mathrm{BA}}^2 - p_2^{\parallel} E_{\parallel}}{3 t_{\mathrm{BA}} p_3^{\perp}}\sqrt{\frac{8x}{m(m-2)}},\\ 
        \text{where} \ m = \frac{3}{4}+\frac{3}{2}x+\frac{1}{4}\sqrt{36x^2-60x+9}\nonumber\\
        \text{and}\ x=\frac{3(e_4-p_4^{\parallel} E_{\parallel})t_{\mathrm{BA}}^2}{2 e_4 t_{\mathrm{BA}}^2+p_2^{\parallel} E_{\parallel}}.\nonumber
    \end{gather}
    Eq.4 approximates Eq.~\ref{eqn:exact_Eperp} to the order of 0.5 of $E_{\parallel}$.

\clearpage

\section{Supplemental figures}

\clearpage

\clearpage

\clearpage

\begin{figure}[h]
    \includegraphics[width=0.6\textwidth]{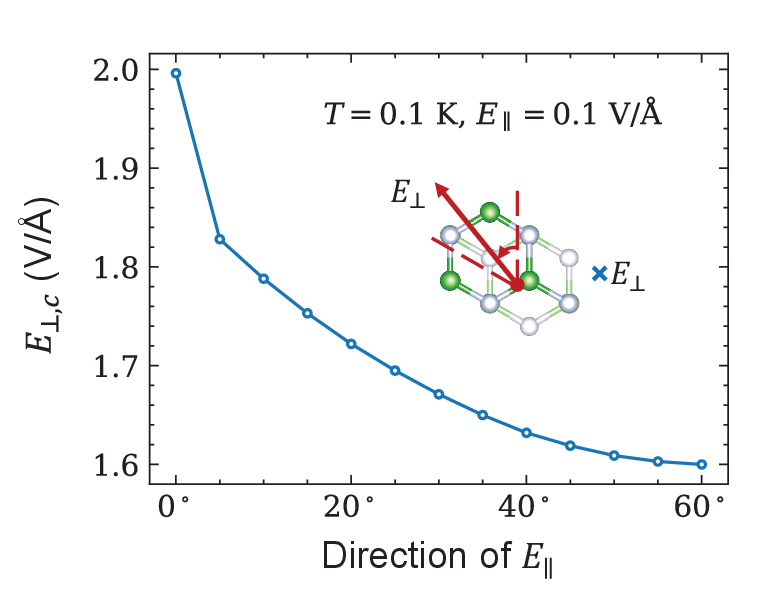}
    \caption{\label{fig:S4-E_random} The variation of the MD simulation obtained $E_{\perp, c}$ with $E_{\parallel}$ orienting from $0^{\circ}$ to $60^{\circ}$ at a fixed magnitude of 0.1V/{\AA} at 0.1K. $E_{\perp, c}$ is determined by identifying the SP state during the sliding process when $E_{\perp}$ along the $-z$ direction is incrementally increasing over time at a rate of 0.02 V/({\AA}$\cdot$ps). Within the range of $\left(0^{\circ}, 60^{\circ} \right]$, the sliding consistently occurs at $60^{\circ}$, while at $0^{\circ}$, the sliding occurs randomly at $60^{\circ}$ or $300^{\circ}$. The closer the direction of $E_{\parallel}$ is to the direction of sliding, the smaller the $E_{\perp, c}$.}
\end{figure}

\clearpage

\begin{figure}[h]
    \includegraphics[width=0.6\textwidth]{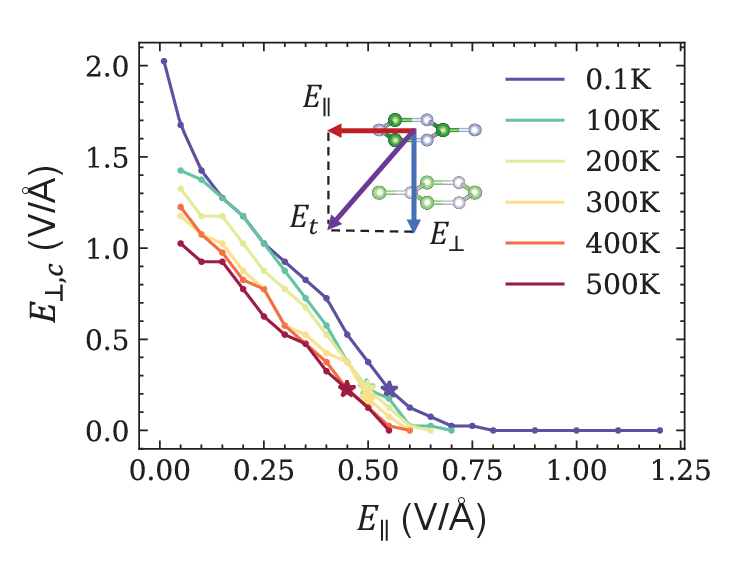}
    \caption{\label{fig:S5-E_slope} The variation of the MD simulation obtained $E_{\perp, c}$ with $E_{\parallel}$ at different magnitudes oriented at $60^{\circ}$ and at different temperatures. $E_{\perp, c}$ is determined as the minimum vertical field, adjusted in increments of 0.05 V/{\AA}, that can initiate sliding within 30 ps. The pentagram marks indicate the the minimum $E_{t, c}$ at each temperature, corresponding to the blue line in Fig.4(c).}
\end{figure}

\clearpage

\begin{figure}[h]
    \includegraphics[width=0.9\textwidth]{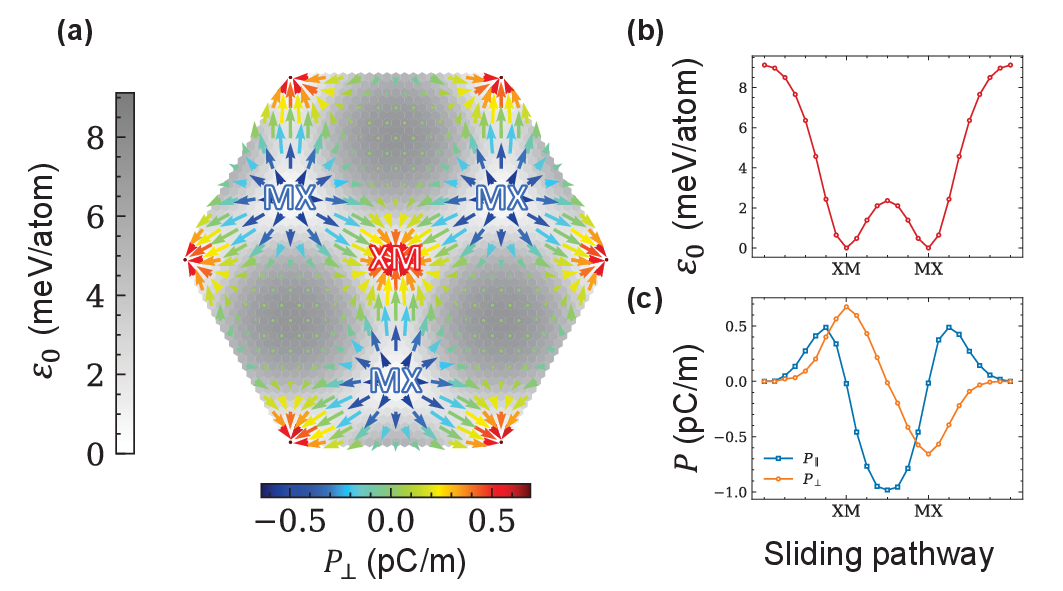}
    \caption{\label{figS6_MoS2} (a) Energy $\varepsilon_0$ and polarization $P$ distribution of $\mathrm{MoS_2}$, calculated by the DFT and Berry phase approach. Positions within the hexagon correspond to different sliding vectors, with the center representing the ferroelectric XM state \cite{WangXirui2022, Weston2020}. The shades of grey in the background indicate energy levels. The direction of colored arrows shows in-plane polarizations $P_{\parallel}$, while their colors represent out-of-plane polarization $P_{\perp}$. Energy $\varepsilon_0$ (b) and polarization $P$ (c) along the designed sliding pathway.}
\end{figure}

\end{document}